\documentclass[sigconf]{acmart}

\AtBeginDocument{%
  \providecommand\BibTeX{{%
    \normalfont B\kern-0.5em{\scshape i\kern-0.25em b}\kern-0.8em\TeX}}}

\copyrightyear{2024}
\acmYear{2024}
\setcopyright{acmlicensed}
\acmConference[WWW '24]{Proceedings of the ACM Web Conference 2024}{May 13--17, 2024}{Singapore, Singapore}
\acmBooktitle{Proceedings of the ACM Web Conference 2024 (WWW '24), May 13--17, 2024, Singapore, Singapore}
\acmDOI{10.1145/3589334.3645467}
\acmISBN{979-8-4007-0171-9/24/05}

\usepackage{booktabs}
\usepackage{multirow}
\usepackage{algorithm}
\usepackage{algorithmic}
\usepackage{enumitem}
\usepackage{enumerate}
\usepackage{xcolor}
\usepackage{graphicx}
\usepackage{subcaption}
\usepackage{diagbox}

\newcommand{\eg}{\emph{e.g.}}
\newcommand{\ie}{\emph{i.e.}}

\begin{document}

\title{ReLLa: Retrieval-enhanced Large Language Models for Lifelong \\ Sequential Behavior Comprehension in Recommendation}

\author{Jianghao Lin}
\email{chiangel@sjtu.edu.cn}
\affiliation{
  \institution{Shanghai Jiao Tong University}
  \city{Shanghai}
  \country{China}
}
\author{Rong Shan}
\email{shanrong@sjtu.edu.cn}
\affiliation{
  \institution{Shanghai Jiao Tong University}
  \city{Shanghai}
  \country{China}
}
\author{Chenxu Zhu}
\email{zhuchenxu1@huawei.com}
\affiliation{
  \institution{Huawei Noah's Ark Lab}
  \city{Shenzhen}
  \country{China}
}
\author{Kounianhua Du}
\email{774581965@sjtu.edu.cn}
\affiliation{
  \institution{Shanghai Jiao Tong University}
  \city{Shanghai}
  \country{China}
}
\author{Bo Chen}
\email{chenbo116@huawei.com}
\affiliation{
  \institution{Huawei Noah's Ark Lab}
  \city{Shenzhen}
  \country{China}
}
\author{Shigang Quan}
\email{quan123@sjtu.edu.cn}
\affiliation{
  \institution{Shanghai Jiao Tong University}
  \city{Shanghai}
  \country{China}
}
\author{Ruiming Tang}
\email{tangruiming@huawei.com}
\affiliation{
  \institution{Huawei Noah's Ark Lab}
  \city{Shenzhen}
  \country{China}
}
\author{Yong Yu}
\email{yyu@sjtu.edu.cn}
\affiliation{
  \institution{Shanghai Jiao Tong University}
  \city{Shanghai}
  \country{China}
}
\author{Weinan Zhang}
\authornote{Weinan Zhang is the corresponding author.}
\email{wnzhang@sjtu.edu.cn}
\affiliation{
  \institution{Shanghai Jiao Tong University}
  \city{Shanghai}
  \country{China}
}

\renewcommand{\shortauthors}{Jianghao Lin et al.}

\begin{abstract}

With large language models (LLMs) achieving remarkable breakthroughs in natural language processing (NLP) domains, LLM-enhanced recommender systems have received much attention and have been actively explored currently. 
In this paper, we focus on adapting and empowering a pure large language model for zero-shot and few-shot recommendation tasks.
First and foremost, we identify and formulate the \emph{lifelong sequential behavior incomprehension problem} for LLMs in recommendation domains, \ie, LLMs fail to extract useful information from a textual context of long user behavior sequence, even if the length of context is far from reaching the context limitation of LLMs. 
To address such an issue and improve the recommendation performance of LLMs, we propose a novel framework, namely \underline{\textbf{R}}etrieval-\underline{\textbf{e}}nhanced \underline{\textbf{L}}arge \underline{\textbf{La}}nguage models (ReLLa) for recommendation tasks in both zero-shot and few-shot settings. 
For zero-shot recommendation, we perform semantic user behavior retrieval (SUBR) to improve the data quality of testing samples, which greatly reduces the difficulty for LLMs to extract the essential knowledge from user behavior sequences.
As for few-shot recommendation, we further design retrieval-enhanced instruction tuning (ReiT) by adopting SUBR as a data augmentation technique for training samples. 
Specifically, we develop a mixed training dataset consisting of both the original data samples and their retrieval-enhanced counterparts.
We conduct extensive experiments on three real-world public datasets to demonstrate the superiority of ReLLa compared with existing baseline models, as well as its capability for lifelong sequential behavior comprehension. 
\textbf{To be highlighted, with only less than 10\% training samples, \emph{few-shot} ReLLa can outperform traditional CTR models that are trained on the entire training set (\eg, DCNv2, DIN, SIM).}
The code is available\footnote{PyTorch version: \url{https://github.com/LaVieEnRose365/ReLLa}}\footnote{MindSpore version: \url{https://github.com/mindspore-lab/models/tree/master/research/huawei-noah/ReLLa}}.

\end{abstract}

\ccsdesc[500]{Information systems~Recommender systems}

\keywords{Large Language Models; Recommender Systems; User Modeling}

\maketitle

\section{Introduction}

\begin{figure}[t]
  \centering
  \includegraphics[width=0.45\textwidth]{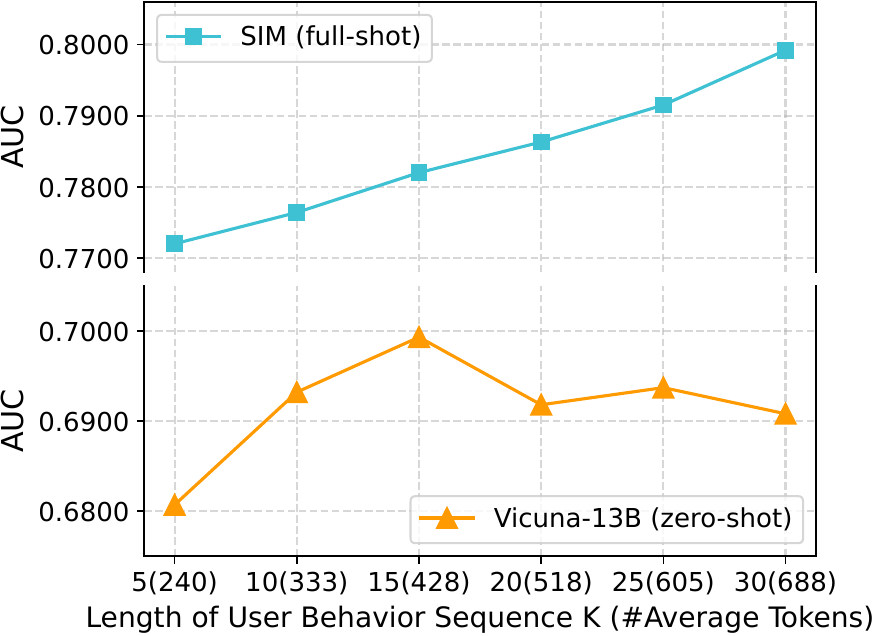}
  \vspace{-5pt}
  \caption{The illustration of lifelong sequential behavior incomprehension problem for LLMs. We report the AUC performance of SIM and Vicuna-13B on MovieLens-1M dataset. 
  While SIM enjoys steady performance improvement as the length of behavior sequence $K$ grows, Vicuna-13B only peaks at $K=15$ and fails to extract the useful information with further longer sequences (\ie, $K>15$).}
  \vspace{-5pt}
  \label{fig:illustration long context problem}
\end{figure}

Recommender systems play a vital role in various online applications to alleviate the information overload problem and satisfy the users' information needs~\cite{DeepFM,xi2023bird,xi2023towards}.
Besides, large language models (LLMs) have flourished in the natural language processing (NLP) domain, showing impressive capacities in generating human-like texts for a wide range of tasks~\cite{brown2020language,touvron2023llama,wang2023recmind,zhang2023memory}.
Consequently, recent works start to explore the potential of LLMs for recommender systems~\cite{lin2023can,hou2023large,bao2023tallrec}. They adopt LLMs directly for various recommendation tasks (\eg, listwise ranking, pointwise scoring), and find out that large language models depict promising performance in zero-shot and few-shot settings for recommendation~\cite{zhang2023recommendation,bao2023tallrec}.

In this paper, we focus on adapting and empowering a pure large language model for recommendation tasks in zero-shot and few-shot settings. 
First, we identify the \textbf{lifelong sequential behavior incomprehension problem}, \ie, \emph{LLMs fail to extract the useful information from a textual context of long user behavior sequence for recommendation tasks, even if the length of context is far from reaching the context limitation of LLMs}. This problem is shown in Figure~\ref{fig:illustration long context problem}, where Vicuna-13B~\cite{vicuna2023, touvron2023llama} is a popular open-source large language model with a context window of 2048 tokens.
As we can observe, the traditional recommendation model (\ie, SIM) enjoys steady performance gains as the length of involved user sequence $K$ grows. 
However, the performance of Vicuna-13B reaches the peak at length $K=15$ and starts to decrease with longer behavior sequence $K>15$, even if the number of involved tokens is far less than the context window limitation (\ie, 2048 tokens). 
While in common NLP tasks, LLMs can definitely exhibit exceptional performance if given a similar length of context (around 600+ tokens).
Therefore, we argue that such an incomprehension problem on long user behavior sequence is special for LLMs in recommendation domains, where it is a rather difficult reasoning task to infer the user's preference towards a certain candidate item based on the given user profile and behavior history.


To address the lifelong sequential behavior incomprehension problem, we propose a novel framework to develop \underline{\textbf{R}}etrieval-\underline{\textbf{e}}nhanced \underline{\textbf{L}}arge \underline{\textbf{La}}nguage models (ReLLa) for recommendation tasks in both zero-shot and few-shot settings. 
For \emph{zero-shot recommendation}, we propose to conduct semantic user behavior retrieval (SUBR) to replace the simply truncated top-$K$ recent behaviors with the top-$K$ semantically relevant behaviors towards the target item. 
In this way, we improve the quality of data samples and reduce the difficulty for LLMs to extract useful information from user behavior sequences, therefore alleviating the incomprehension problem.
For \emph{few-shot recommendation}, apart from applying SUBR to improve the data quality of samples, we propose to perform retrieval-enhanced instruction tuning (ReiT) to further promote the ability of LLMs to handle inputs with long behavior sequences. 
We apply SUBR on training samples as the data augmentation techniques to obtain a mixed training dataset of both original and retrieval-enhanced training data samples, which increases the robustness and generalization ability of LLMs.
\textbf{More surprisingly, with only few-shot training samples (\eg, 8,192 data instances in MovieLens-25M dataset), ReLLa can outperform full-shot traditional recommendation models (\eg, DCNv2~\cite{DCNv2}, DIN~\cite{zhou2018deep}, and  SIM~\cite{SIM}) that are trained with the entire training set (\eg, nearly 20M samples in MovieLens-25M dataset}).

Main contributions of this paper are in three folds:
\begin{itemize}[leftmargin=10pt]
    \item To the best of our knowledge, we are the first to identify and well formulate the lifelong sequential behavior incomprehension problem for LLMs in recommendation, where LLMs are generally incomprehensible to a textual context of long user behavior sequence, even if the length of context is far from reaching the context limitation.
    \item We propose a novel ReLLa (\underline{\textbf{R}}etrieval-\underline{\textbf{e}}nhanced \underline{\textbf{L}}arge \underline{\textbf{La}}nguage Models) framework to mitigate the incomprehension problem of LLMs on long user behavior sequences. 
    We design semantic user behavior retrieval (SUBR) to improve the data quality of data samples for zero-shot recommendation, and further propose retrieval-enhanced instruction tuning (ReiT) to promote the few-shot recommendation performance with a mixture of original and retrieval-enhanced training samples.
    \item Extensive experiments on three real-world public datasets validate the effectiveness of our method compared with existing baselines. 
    \textbf{Note that the baseline models are trained in \emph{full-shot} settings with the entire training set, while ReLLa is only trained with \emph{few-shot} samples.}
\end{itemize}
\section{Preliminaries}
\label{sec:preliminary}

In this paper, we focus on the click-through rate (CTR) prediction, which serves as the core component in recommender systems to estimate a user's click probability towards a target item given a certain context~\cite{lin2023map,DeepFM}. 
The training dataset for CTR prediction is denoted as $\{(x_i,y_i)\}_{i=1}^N$, where $N$ is the number of data samples (\ie, $N$-shot).
When adapting a pure large language model for such a pointwise scoring task, we need to clarify the following three key aspects: (1) what is the definition of zero-shot and few-shot recommendations, (2) how to formulate the textual input-output pairs, and (3) how to do pointwise scoring with LLMs. 

\subsection{Zero-shot and Few-shot Recommendations}

Zero-shot recommendation implies that a model is directly employed for the target recommendation task without any tuning on the in-domain training data. 
Apparently, traditional recommendation models are incapable of accomplishing zero-shot recommendation tasks, since they are randomly initialized.
However, LLMs possess a vast volume of open-world knowledge and logical reasoning abilities, which enable them to infer the user's preference towards a certain target item based on the profile of user/item. 

Few-shot recommendation refers to low-resource scenarios with $N$ training data samples. $N$ denotes the number of shots, which is a relatively small number. 
This highly requires the data efficiency characteristic of an algorithm to fully exploit the limited number of training samples to achieve better recommendation performance.

Extending from the definition of few-shot recommendation, we can therefore define full-shot recommendation as the setting where we train the model based on the entire training set.

\subsection{Textual Input-Output Pair Formulation}\label{sec:Textual Input-Output Pair Formulation}

\begin{figure}[h]
  \centering
  \vspace{-5pt}
  \includegraphics[width=0.46\textwidth]{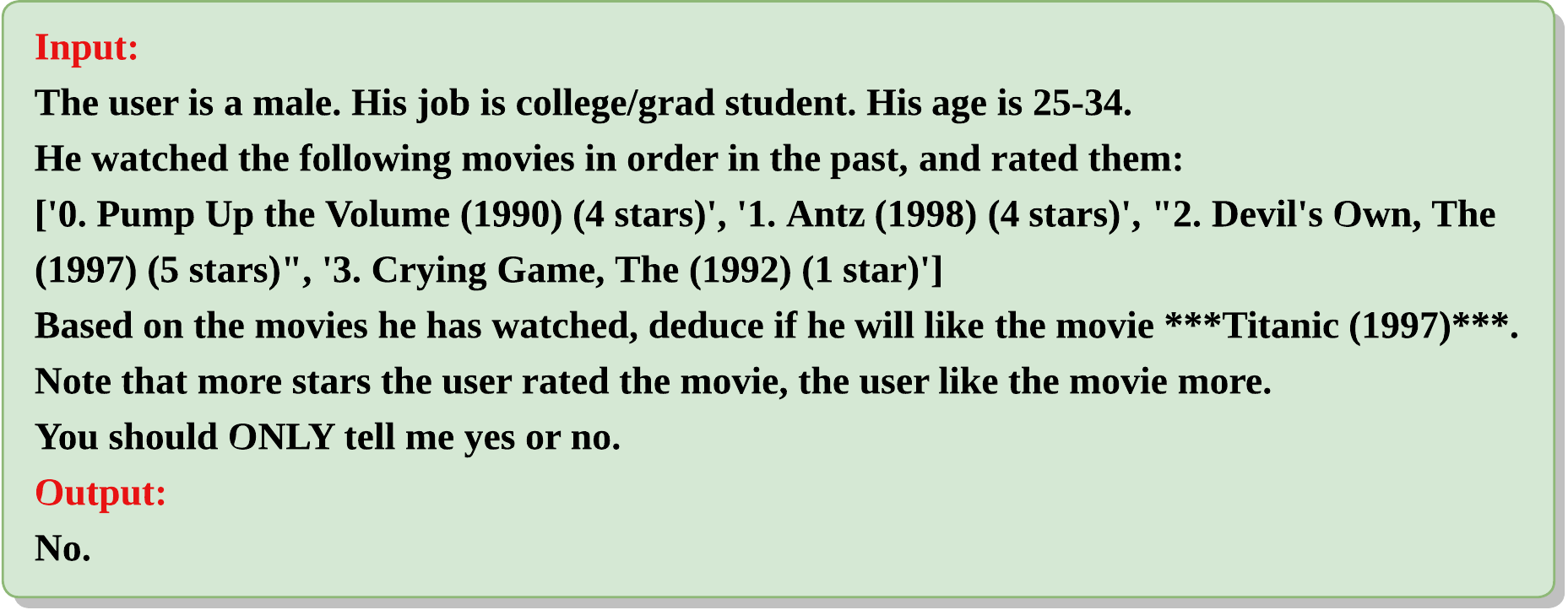}
  \vspace{-5pt}
  \caption{Illustration of textual input-output pair.}
  \vspace{-10pt}
  \label{fig:input output pair example}
\end{figure}

For LLMs, we need to convert each data sample $x_i$ into textual sentences $x_i^{text}$ via hard prompt templates. 
Similarly, the binary label $y_i\in\{0,1\}$ is transformed into a pair of binary key answer words $y_i^{text}\in\{\text{``Yes''}, \text{``No''}\}$. We give an illustrative example of the input-output pair $(x_i^{text},y_i^{text})$ in Figure~\ref{fig:input output pair example}, where $x_i^{text}$ contains the descriptive texts for user profile, user behavior sequence, target item and task description, respectively. 

Notably, the predominant factor that determines the length of context is derived from the user behavior sequence, the length of which can varies from tens to hundreds. 
For each input $x_i$, we truncate the user behavior sequence to length $K$. 
For example, the length of behavior sequence in Figure~\ref{fig:input output pair example} is $K=4$. 
While the common sequential CTR prediction settings usually truncate and adopt \emph{the most recent $K$ behaviors}, ReLLa propose to conduct semantic user behavior retrieval to construct textual inputs with \emph{the most relevant $K$ behaviors} towards the target item.

\subsection{Pointwise Scoring with LLMs}
\label{sec:scoring with LLM}

The large language model takes as input the discrete tokens of $x_i^{text}$, and generate the next token $\hat{y}_i^{text}$ as the output, the process of which can be formulated as follows:
\begin{equation}
\begin{aligned}
    s_i &= \operatorname{LLM}(x_i^{text})\,\in\mathbb{R}^{V}, \\
    p_i &= \operatorname{Softmax}(s_i)\,\in\mathbb{R}^V,\\
    \hat{y}_i^{text} &\sim p_i \,,
\end{aligned}
\end{equation}
where $V$ is the vocabulary size, and $\hat{y}_i^{text}$ is the next predicted token sampled from the probability distribution $p_i$. 

However, CTR prediction requires the model to do pointwise scoring, and the output should be floating-point number $\hat{y}_i\in[0,1]$ instead of a discrete token $\hat{y}_i^{text}$.
Therefore, following previous works~\cite{bao2023tallrec,zhang2023prompt}, we intercept the estimated scores $s_i\in\mathbb{R}^V$, and conduct a bidimensional softmax over the corresponding scores of the binary key answer words. 
Suppose the vocabulary indices for ``Yes'' and ``No'' are $a$ and $b$, respectively. 
The pointwise scoring of LLMs for CTR prediction can be written as:
\begin{equation}
\begin{aligned}
    \hat{y}_i = \frac{\exp(s_{i,a})}{\exp(s_{i,a})+\exp(s_{i,b})} \,\in(0,1).
\end{aligned}
\end{equation}

It is worth noting that such an estimated click-through rate $\hat{y}_i$ is only leveraged for evaluation on the testing set. 
We preserve the common instruction tuning and causal language modeling paradigm for LLMs if training is involved.
\section{Methodology}
\label{sec:method}

\begin{figure}[t]
  \centering
  
  \vspace{-7pt}
  \includegraphics[width=0.45\textwidth]{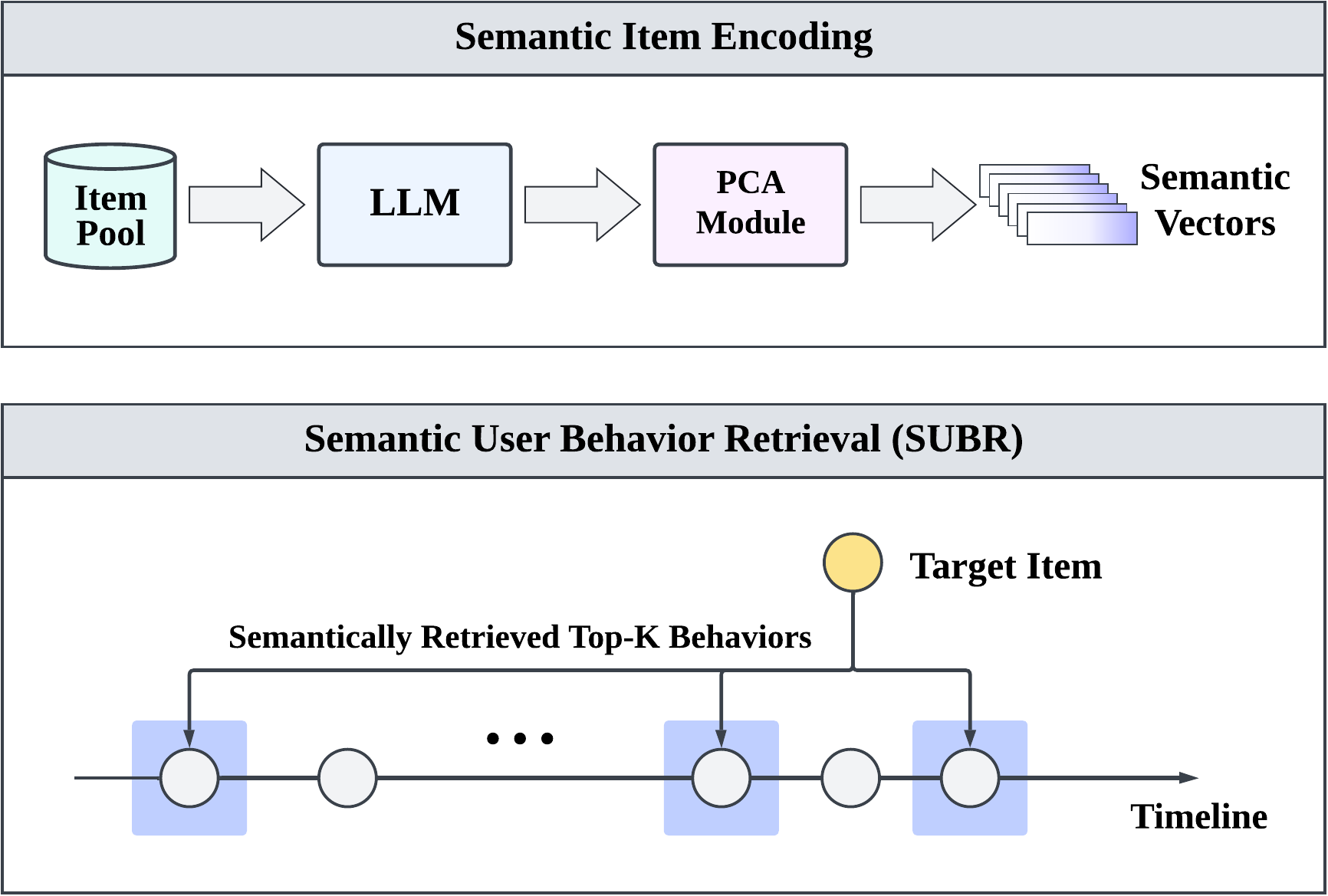}
  \vspace{-5pt}
  \caption{Illustration of semantic user behavior retrieval (SUBR), which improves the data quality by retrieving the top-$K$ semantically relevant behaviors towards the target item. 
  This reduces the difficulty for LLMs to extract useful information from the user history, and therefore alleviates the long user behavior sequence incomprehension problem.}
  \vspace{-15pt}

  \label{fig:semantic user behavior retrieval}
\end{figure}

In this section, we introduce the ReLLa (\underline{\textbf{R}}etrieval-\underline{\textbf{e}}nhanced \underline{\textbf{L}}arge \underline{\textbf{La}}nguage Models) framework in details. 

\subsection{Overview of ReLLa}
In the ReLLa framework, we develop two key techniques for LLMs in zero-shot and few-shot recommendations, respectively. 

For zero-shot recommendation, as illustrated in Figure~\ref{fig:semantic user behavior retrieval}, we propose to conduct semantic user behavior retrieval (SUBR) to improve the data quality of data samples.
We first leverage the large language model to obtain the semantic vectors for each item.
Then, for each textual data sample $x_i^{text}$, we retrieve the most \emph{semantically relevant} $K$ behaviors, which can substitute the original most \emph{recent} $K$ behaviors.

For few-shot recommendation, as shown in Figure~\ref{fig:retrieval-enhanced instruction tuning}, we propose to perform retrieval-enhanced instruction tuning (ReiT) to promote the ability of LLMs to extract useful information from long behavior sequences. 
Notably, the semantic user behavior retrieval (SUBR) is adopted as the data augmentation technique to form the mixed training dataset. 
The mixture of both original and retrieval-enhanced data samples introduces more variety and patterns in the training set, thus increasing the robustness and generalization ability of LLMs for lifelong sequential behavior comprehension.

Although ReLLa is tuned in \textbf{\emph{few-shot}} settings, we would like to again emphasize that other recommendation baseline models are trained in \textbf{\emph{full-shot}} settings with the entire training set.

\subsection{Semantic User Behavior Retrieval}
\label{sec:semantic UBR}

In zero-shot settings, the parameters of LLMs cannot be tuned according to the in-domain training samples.
Hence, as shown in Figure~\ref{fig:semantic user behavior retrieval}, semantic user behavior retrieval (SUBR) aims to improve the quality of each sample by replacing the simply truncated most recent $K$ behaviors with the most semantically relevant $K$ behaviors towards the target item. 
As suggested in previous works~\cite{qin2020user,SIM}, the retrieved user behaviors can denoise the user history and convey more clear and essential user interests for the target item, while preserving the original length of user sequence as the model input.

\begin{figure}[h]
  \centering
  \vspace{-3pt}
  \includegraphics[width=0.47\textwidth]{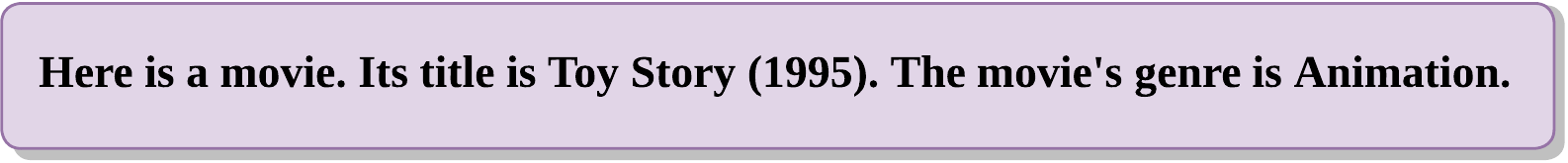}
  \vspace{-5pt}
  \caption{Illustration of descriptive text for an item (movie).}
  \vspace{-5pt}
  \label{fig:item description}
\end{figure}

Firstly, we conduct semantic item encoding to obtain the semantic vector for each item. 
For the $t$-th item in the pool, a descriptive text is constructed via hard prompt template (an example is given in Figure~\ref{fig:item description}, and is then fed through LLM. 
We perform average pooling over all the hidden states from the last layer of LLM, resulting in a vector $u_t\in\mathbb{R}^D$, where $D$ is the hidden size of LLM (\eg, 4096 for Vicuna-7B, and 5120 for Vicuna-13B).
A principal component analysis (PCA)~\cite{shlens2014tutorial} module is further employed for both dimension reduction and denoising purposes, engendering the final semantic representation $v_t\in\mathbb{R}^d$, where we set $d=512$. 
Now we can measure the semantic relevance between each pair of items via the cosine similarity between their corresponding semantic representations.

Next, we can apply semantic user behavior retrieval on each testing sample to replace the original truncated top-$K$ recent behaviors with the top-$K$ semantically relevant behaviors towards the target item. 
In this way, we obtain a parallel retrieval-enhanced testing dataset with higher data quality, while keeping the length of input context roughly unchanged. 
Therefore, SUBR can improve the zero-shot recommendation performance, and mitigate the incomprehension problem on long user behavior sequences.

\subsection{Retrieval-enhanced Instruction Tuning}

\begin{figure}[t]
  \centering
  \vspace{-8pt}
  \includegraphics[width=0.46\textwidth]{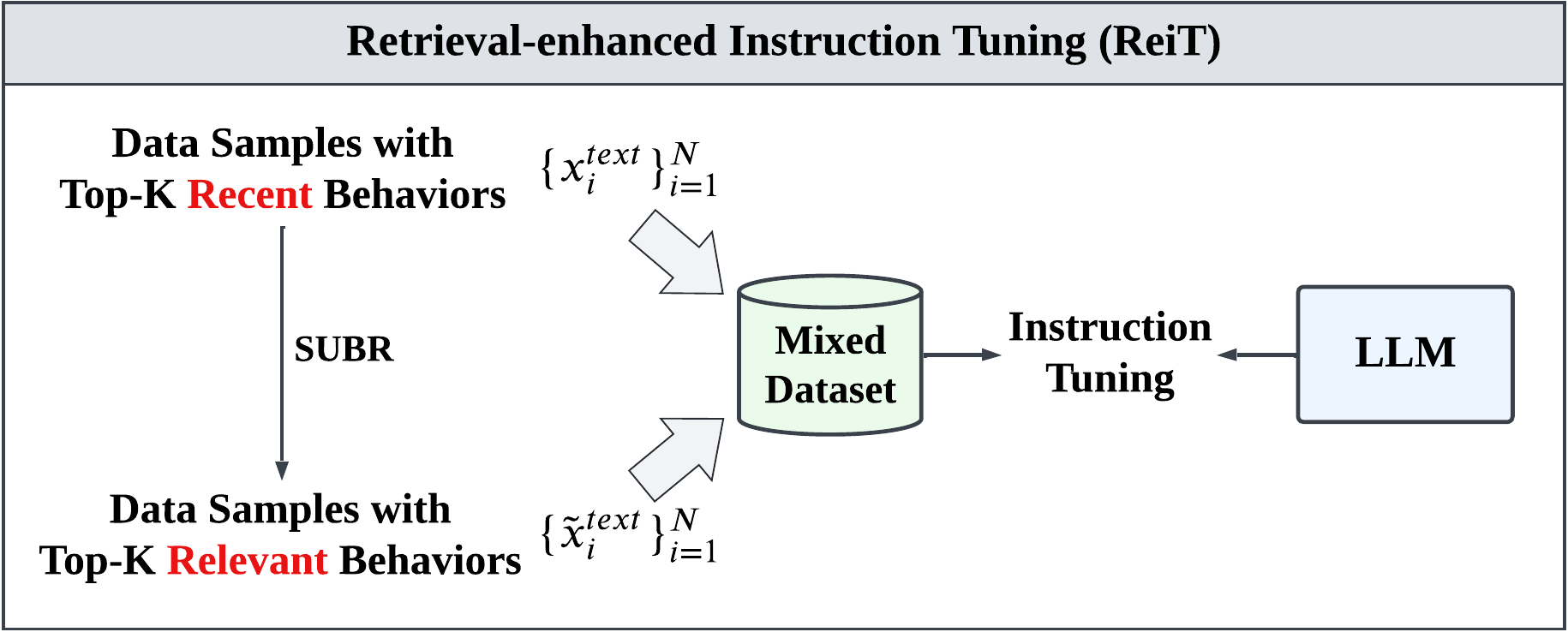}
  \vspace{-5pt}
  \caption{Illustration of retrieval-enhanced instruction tuning, where we construct a mixed training dataset.
  The mixed dataset consists of both the original textual input-output samples and their retrieval-enhanced counterparts obtained via semantic user behavior retrieval (SUBR).}
  \vspace{-10pt}
  \label{fig:retrieval-enhanced instruction tuning}
\end{figure}

As for few-shot recommendation, we denote the training dataset as $\{(x_i^{text},y_i^{text})\}_{i=1}^N$, where $N$ is the number of shots (\ie, training samples). 
While previous works~\cite{bao2023tallrec,zhang2023recommendation} directly employ instruction tuning for LLMs over the converted textual input-output pairs, we argue that simple instruction tuning could potentially expose large language models to risks of overfitting and catastrophic forgetting on limited number of training data~\cite{korbak2022controlling,ramasesh2021effect}.

To this end, we propose a novel retrieval-enhanced instruction tuning (ReiT), where semantic user behavior retrieval (SUBR) is adopted as the data augmentation technique to construct a mixed training dataset with enriched user behavior patterns.
As shown in Figure~\ref{fig:retrieval-enhanced instruction tuning}, we apply SUBR on each training data to obtain its retrieval-enhanced counterpart
$\tilde{x}_i^{text}$. 
Next, we merge the original and retrieval-enhanced data instances to construct a mixed training dataset with total $2N$ samples. Finally, we conduct instruction tuning for LLMs on the mixed training data.
The pattern enrichment brought by SUBR can regularize and prevent the large language model from overfitting, thus promoting its robustness and generalization ability to effectively extract essential knowledge from a long user behavior sequence of length $K$.

We leverage the causal language modeling objective for instruction tuning to retain the original model structure:
\begin{equation}
    \max_{\Theta}\sum\nolimits_{(x,y)\in\mathcal{M}}\, \sum\nolimits_{j=1}^{|y|}\log P_{\Theta}(y_{j}|x,y_{<j}),
\end{equation}
where $\Theta$ is the parameter of LLM, $\mathcal{M}$ is the mixed training dataset with total $2N$ data samples, $y_j$ is the $j$-th token of the textual output $y$, and $y_{<j}$ denotes the tokens before $y_j$.
There is \emph{no} randomly initialized prediction layer appended upon LLM for CTR prediction with binary cross-entropy (BCE) loss. 
The CTR estimation method for pointwise scoring with LLMs discussed in Section~\ref{sec:scoring with LLM} is only used for evaluation on the testing set.

While we maintain a mixed training dataset for instruction tuning, the testing set contains pure retrieval-enhanced data samples generated by SUBR, which is the same as zero-shot recommendation as described in Section~\ref{sec:semantic UBR}.
Moreover, we provide further discussion about ReiT to address readers' possible concerns as follows:
\begin{itemize}[leftmargin=10pt]
    \item \emph{Will ReiT cause the inconsistency between the training and testing data?} Data augmentation is a common regularization technique, especially for low-resource few-shot settings in computer vision (CV)~\cite{berthelot2019mixmatch,zhang2017mixup} or natural language processing (NLP)~\cite{li2022data,feng2021survey}. 
    The inconsistency would not exist, as long as the augmentation algorithm is sound and reasonable. 
    \item 
    \emph{Which factor actually contribute to the performance improvement of ReiT? The doubled training samples, or the pattern enrichment?} 
    Both factors can lead to the final performance enhancement, but we argue that the pattern enrichment as regularization is a more important factor for model robustness. 
    Empirical studies are provided in Section~\ref{sec:ablation} to ablate and decouple these two factors.
\end{itemize}

\section{Experiment}

In this section, we conduct extensive experiments to answer the following research questions:
\begin{itemize}
    \item[\textbf{RQ1}] How does ReLLa perform compared to existing baselines?
    \item[\textbf{RQ2}] Does ReLLa promote the lifelong sequential behavior comprehension ability of LLMs for recommendation tasks?
    \item[\textbf{RQ3}] How does the number of shots $N$ affect the performance?
    \item[\textbf{RQ4}] What are the influences of different components for ReLLa?
    \item[\textbf{RQ5}] How ReLLa help LLMs to better comprehend the user behavior sequence?
\end{itemize}

Due to the page limitation, we further provide additional experiments in Appendix~\ref{app:exp} to verify the following core points:
\begin{itemize}[leftmargin=10pt]
    \item The \textit{universality} of the lifelong sequential behavior incomprehension problem and the \textit{generalization} of our proposed ReLLa.
    \item Analysis about the model parameter and inference time.
    \item Ablation on PCA dimensionality and distance metrics for SUBR.
    \item Analysis of potential reasons for the incomprehension problem.
\end{itemize}

\subsection{Experiment Setup}
\subsubsection{Datasets}
We conduct experiments on three real-world datasets (\ie, BookCrossing\footnote{\url{http://www2.informatik.uni-freiburg.de/~cziegler/BX/}}, MovieLens-1M\footnote{\url{https://grouplens.org/datasets/movielens/1m/}} and MovieLens-25M\footnote{\url{https://grouplens.org/datasets/movielens/25m/}}). 
We show the dataset statistics in Table~\ref{tab:datasets} and give detailed data preprocessing information in Appendix~\ref{app:dataset} due to page limitations.


\begin{table}
    \vspace{-5pt}
    \caption{The dataset statistics.}
    \vspace{-5pt}
    \centering
    \resizebox{0.45\textwidth}{!}{
    \renewcommand\arraystretch{1.1}
    \begin{tabular}{c|cccccc}
    \toprule
     Dataset   & \#Users & \#Items & \#Samples & \#Fields & \#Features \\ 
     \midrule
     BookCrossing  & 278,858 & 271,375 & 17,714 & 10 & 912,279 \\
     MovieLens-1M & 6,040 & 3,706 & 970,009 & 10 & 16,944 \\
     MovieLens-25M & 162,541 & 59,047 & 25,000,095 & 6 & 280,576 \\ \bottomrule
    \end{tabular}
    }
    \vspace{-5pt}
    \label{tab:datasets}
\end{table}




\subsubsection{Evaluation Metrics}
To evaluate the performance of our methods, we leverage AUC (area under the ROC curve), Log Loss (binary cross-entropy loss) and ACC (accuracy score) as the evaluation metrics. 
In CTR prediction, slightly higher AUC or lower Log Loss (e.g., 0.001) can be regarded as significant improvement~\cite{xDeepFM, DCNv2}.

\subsubsection{Baseline Models}
The CTR baseline models can be mainly classified into two categories: (1) \emph{traditional CTR models} that take one-hot encoded IDs as inputs, and (2) \emph{LM-based models} that incorporate pretrained language models and formulate CTR prediction as either text classification or sequence-to-sequence problem.

Traditional CTR models can be further categorized into (1) feature interaction models, and (2) user behavior models.
We select DeepFM~\cite{DeepFM}, AutoInt~\cite{AutoInt}, and DCNv2~\cite{DCNv2} as representative feature interaction models, and choose GRU4Rec~\cite{GRU4Rec}, Caser~\cite{Caser}, SASRec~\cite{SASRec}, DIN~\cite{zhou2018deep}, and SIM~\cite{SIM} as representative user behavior models. 
We apply average pooling over users' historical behaviors, and regard the outputs as additional feature fields for the feature interaction models.
SIM~\cite{SIM} is a classical sequential CTR model that leverages user behavior retrieval techniques to enhance the recommendation performance. We include it for fair comparison, since ReLLa incorporates semantic user behavior retrieval (SUBR).
As for LM-based CTR models, we select CTR-BERT~\cite{CTRBERT}, PTab~\cite{liu2022ptab}, and P5~\cite{P5} as the representative baselines. 
TALLRec~\cite{bao2023tallrec} adopts the simple instruction tuning framework for LLMs, and we therefore include it in our ablation study in Section~\ref{sec:ablation}.


\begin{table*}
\vspace{-7pt}
\caption{The performance of different models in \emph{zero-shot}, \emph{full-shot} and \emph{few-shot} settings. 
In \emph{full-shot} setting, the baselines are trained on the entire training set. 
In \emph{few-shot} setting, the number of training shots $N$ is selected from $\{256 (<1\%), 1024(<10\%)\}$ on BookCrossing dataset, and $\{8192 (<1\%), 65536 (<10\%)\}$ on MovieLens-1M and MovieLens-25M datasets. 
The best result is given in bold, and the second-best value is underlined. 
\emph{Rel.Impr} denotes the relative AUC improvement rate of ReLLa against each baseline. 
The symbol $\ast$ indicates statistically significant improvement of ReLLa over the best baseline with $p$-value < 0.001.
}
\vspace{-5pt}
\label{tab:zero & few shot performance}
\resizebox{0.965\textwidth}{!}{
\renewcommand\arraystretch{1.1}
\begin{tabular}{c|c|cccc|cccc|cccc}
\toprule
\hline

\multicolumn{2}{c|}{\multirow{2}{*}{Model}} & \multicolumn{4}{c|}{BookCrossing} & \multicolumn{4}{c|}{MovieLens-1M} & \multicolumn{4}{c}{MovieLens-25M} \\ 
\multicolumn{2}{c|}{} & AUC  & Log Loss & ACC & Rel.Impr & AUC  & Log Loss & ACC & Rel.Impr & AUC  & Log Loss & ACC & Rel.Impr\\ 
   \hline 
   
\multicolumn{1}{c|}{\multirow{3}{*}{Zero-shot}} & Vicuna-7B & 0.7011 & \underline{0.9357} & 0.5378 & 3.45\% & 0.6739 & 0.9510 & 0.5644 & 4.07\% & \underline{0.7468} & 0.6348 & 0.6392 & -1.93\% \\
\multicolumn{1}{c|}{\multirow{3}{*}{}} & Vicuna-13B & \underline{0.7176} & 0.9507 & \underline{0.5649} & 1.07\% & 0.6993 & 0.6291 & 0.6493 & 0.29\% & \textbf{0.7503} & \underline{0.6308} & \underline{0.6427} & -2.39\% \\
\multicolumn{1}{c|}{\multirow{3}{*}{}} & ReLLa (Ours) & \textbf{0.7253$^*$} & \textbf{0.9277$^*$} & \textbf{0.5750$^*$} & - & \textbf{0.7013$^*$} & \textbf{0.6250$^*$} & \textbf{0.6507$^*$} & - & 0.7324 & \textbf{0.5858$^*$} & \textbf{0.7027$^*$} & - \\
   \hline

\multicolumn{1}{c|}{\multirow{12}{*}{Full-shot}} & DeepFM & 0.7496 & 0.5953 & 0.6760 & 1.05\% & 0.7915 & 0.5484 & 0.7225 & 1.49\% & 0.8189 & 0.4867 & 0.7709 & 3.52\% \\ 
\multicolumn{1}{c|}{\multirow{4}{*}{}} & AutoInt & 0.7481 & 0.6840 & 0.6365 & 1.26\% & 0.7929 & 0.5453 & 0.7226 & 1.31\% & 0.8169 & 0.4957 & 0.7689 & 3.77\% \\ 
\multicolumn{1}{c|}{\multirow{4}{*}{}} & DCNv2 & 0.7472 & 0.6816 & 0.6472 & 1.38\% & 0.7931 & 0.5464 & 0.7216 & 1.29\% & 0.8190 & 0.4989 & 0.7702 & 3.50\%\\ 
\multicolumn{1}{c|}{\multirow{4}{*}{}} & GRU4Rec & 0.7479 & 0.5930 & 0.6777 & 1.28\% & 0.7926 & 0.5453 & 0.7225 & 1.35\% & 0.8186 & 0.4941 & 0.7700 & 3.55\% \\ 
\multicolumn{1}{c|}{\multirow{4}{*}{}} & Caser & 0.7478 & 0.5990 & 0.6760 & 1.30\% & 0.7918 & 0.5464 & 0.7206 & 1.45\% & 0.8199 & 0.4865 & 0.7707 & 3.39\% \\ 
\multicolumn{1}{c|}{\multirow{4}{*}{}} & SASRec & 0.7482 & 0.5934 & \textbf{0.6811} & 1.24\% & 0.7934 & 0.5460 & 0.7233 & 1.25\% & 0.8187 & 0.4956 & 0.7691 & 3.54\% \\ 
\multicolumn{1}{c|}{\multirow{4}{*}{}} & DIN & 0.7477 & 0.6811 & 0.6557 & 1.31\% & 0.7962 & 0.5425 & 0.7252 & 0.89\% & 0.8190 & 0.4906 & 0.7716 & 3.50\% \\ 
\multicolumn{1}{c|}{\multirow{4}{*}{}} & SIM & \underline{0.7541} & \textbf{0.5893} & 0.6777 & 0.45\% & \underline{0.7992} & \underline{0.5387} & \underline{0.7268} & 0.51\% & \underline{0.8344} & \underline{0.4724} & \underline{0.7822} & 1.59\% \\ 
\multicolumn{1}{c|}{\multirow{4}{*}{}} & CTR-BERT & 0.7448 & 0.5938 & 0.6704 & 1.71\% & 0.7931 & 0.5457 & 0.7233 & 1.29\% & 0.8079 & 0.5044 & 0.7511 & 4.93\% \\ 
\multicolumn{1}{c|}{\multirow{4}{*}{}} & PTab & 0.7429 & 0.6154 & 0.6574 & 1.97\% & 0.7955 & 0.5428 & 0.7240 & 0.98\% & 0.8107 & 0.5022 & 0.7551 & 4.56\% \\ 
\multicolumn{1}{c|}{\multirow{4}{*}{}} & P5 & 0.7438 & 0.6128 & 0.6563 & 1.84\% & 0.7937 & 0.5478 & 0.7190 & 1.21\% & 0.8092 & 0.5030 & 0.7527 & 4.76\% \\ 
\hline
\multicolumn{1}{c|}{\multirow{2}{*}{Few-shot}} & ReLLa (<1\%) & 0.7482 & 0.6265 & 0.6800 & - & 0.7927 & 0.5475 & 0.7196 & - & 0.8352 & 0.4693 & 0.7779 & - \\ 
\multicolumn{1}{c|}{\multirow{2}{*}{}} & ReLLa (<10\%) & \textbf{0.7575$^*$} & \underline{0.5919} & \underline{0.6806} & - & \textbf{0.8033$^*$} & \textbf{0.5362$^*$} & \textbf{0.7280$^*$} & - & \textbf{0.8477$^*$} & \textbf{0.4524$^*$} & \textbf{0.7925$^*$} & - \\ 
  
   \hline  
   \bottomrule          
\end{tabular}
\vspace{-5pt}
}
\end{table*}

\subsubsection{Implementation Details}

We select Vicuna-13B~\cite{vicuna2023} released by FastChat\footnote{\url{https://github.com/lm-sys/FastChat}} as the base LLM for ReLLa.
All the experiments are conducted on V100 GPUs.
For training resource efficiency, 8-bit quantization and low-rank adaption (LoRA)~\cite{hu2021lora} are adopted for parameter-efficient finetuning (PEFT). 
We follow previous works~\cite{bao2023tallrec,leng2023chinese-vicuna} to set the configuration of LoRA, with LoRA rank as 8, LoRA alpha as 16, and LoRA dropout as 0.05.
The LoRA update matrices are applied on the query and value projection matrices of attention blocks. 
During instruction tuning, we adopt AdamW~\cite{adamw} optimizer with weight decay set to 0.
The model is trained with a batch size selected from $\{128, 256\}$. The learning rate is initialized from $\{1\times 10^{-3}, 1.5 \times 10^{-3}\}$ with linear scheduler. On BookCrossing dataset, the maximum training epoch is set to 10, while on MovieLens-1M and MovieLens-25M datasets, the maximum epoch is set to 5. 
The configuration of baselines is in Appendix~\ref{app:baseline}. 
The hard prompt templates for textual input-output pairs and item descriptions for all three datasets are in Appendix~\ref{app:prompt}.



Moreover, when constructing the hard prompt template for ReLLa, we remove all the pure ID fields, \ie, \textit{User ID} and \textit{ISBN} fields 
on BookCrossing dataset, \textit{User ID}, \textit{Movie ID}, and \textit{Zipcode} fields on MovieLens-1M dataset, \textit{User ID} and \textit{Movie ID} fields on MovieLens-25M dataset.
The reason is that LLMs possess limited perceptual abilities for pure ID texts~\cite{lin2023can}.
Other fields are leveraged as user profile or item information in the prompt, as described in Section~\ref{sec:Textual Input-Output Pair Formulation} and Appendix~\ref{app:prompt}.
Note that we \emph{\textbf{do not}} discard any features for other CTR baseline models, \ie, they take all the feature fields and user behavior sequences as inputs.


\subsection{Overall Performance (RQ1)}

We evaluate the performance of ReLLa in comparison to existing baseline models, and report the results in Table~\ref{tab:zero & few shot performance}. 
Note that other recommendation baseline models are all trained in \textbf{\emph{full-shot}} settings with the entire training set. 
We set the length of user behavior sequence $K$ to 60/30/30 for BookCrossing/MovieLens-1M/MovieLens-25M, respectively.


For zero-shot recommendation, we observe that:
\begin{itemize}[leftmargin=10pt]
    \item The performance of Vicuna-7B is notably inferior to its 13B version on all three datasets. It demonstrates that
    a larger LLM possesses more excellent language comprehension and logical reasoning abilities, therefore leading to better zero-shot inference capability for user preference.
    \item ReLLa significantly outperforms Vicuna-13B for all three metrics on BookCrossing and MovieLens-1M datasets. Although the AUC performance of ReLLa degenerates on MovieLens-25M, ReLLa attains significant improvements in terms of pointwise metrics (\ie, Log Loss and ACC).
    such phenomena validate the effectiveness of SUBR in reducing the difficulty for LLMs to extract useful information from user behavior sequences. 
    Also, the AUC degeneration of AUC on MovieLens-25M reveals the potential instability of zero-shot LLMs for recommendation.
\end{itemize}
As for full-shot and few-shot settings, we can draw the following observations from Table~\ref{tab:zero & few shot performance}:
\begin{itemize}[leftmargin=10pt]
    \item SIM achieves the best performance among all the baseline models. 
    SIM applies user behavior retrieval to reduce the noise of user sequences, which is essentially beneficial for CTR prediction.
    Besides, LM-based CTR models (\ie, CTR-BERT, PTab, P5) perform worse than most of the ID-based traditional CTR models, which is consistent with the results reported in~\cite{li2023ctrl,rajput2023recommender}. These LM-based methods only incorporate small language models (\eg, BERT~\cite{berthelot2019mixmatch}, T5~\cite{T5}) for pure text-based recommendation, and therefore result in inferior performance.

    \item ReLLa (few-shot) generally achieves significantly better performance over all the baseline models, except for few cases, which validates the effectiveness of our proposed retrieval-enhanced instruction tuning (ReiT).
    \textbf{It is worth noting that ReLLa only utilizes less than 10\% training samples for finetuning, while other baseline models are trained on the entire training set}, \eg, $N=65,536$ for ReLLa and $N=19,349,912$ for SIM on MovieLens-25M dataset.
    This demonstrates the superior data efficiency of ReLLa for sequential recommendation tasks.
    


\end{itemize}

\begin{figure*}[t]
\centering
\vspace{-5pt}
\includegraphics[width=0.97\textwidth]{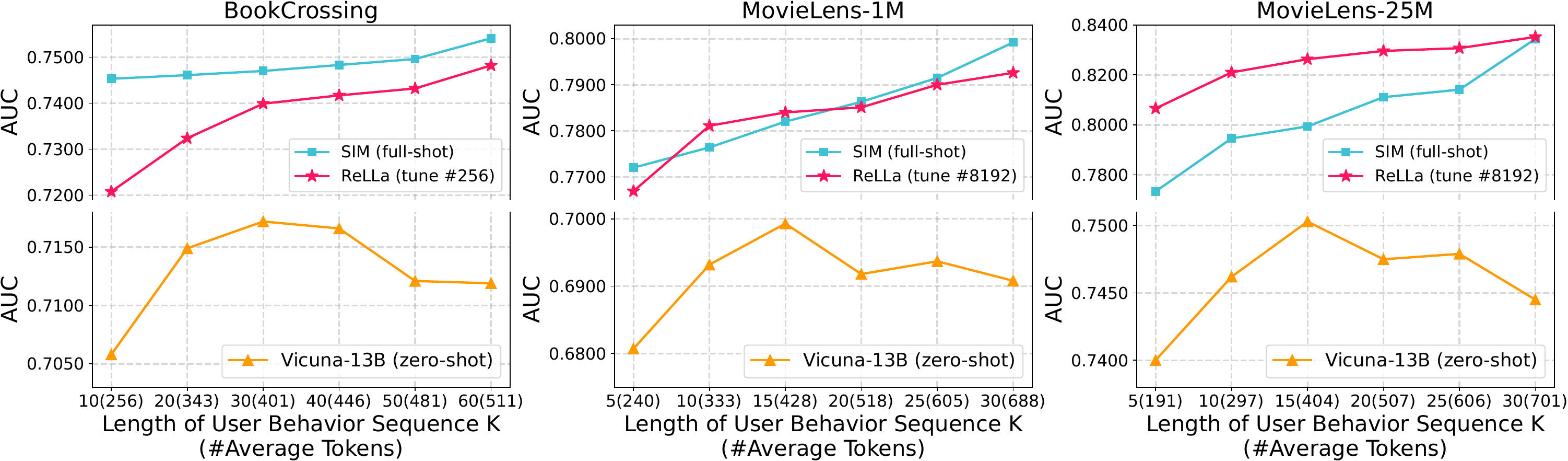}
\vspace{-5pt}
  \caption{The AUC performance of different models w.r.t. different length of user behavior sequence $K$. 
  ReLLa manages to mitigate the incomprehension problem of LLMs on recommendation tasks with long user behavior sequences.
  }
  \vspace{-5pt}
  \label{fig:vary K}
\end{figure*}

\subsection{Sequential Behavior Comprehension (RQ2)}
\label{sec:study on shot K}


We vary the length of user behavior sequence $K$ to investigate its impact on CTR prediction performance, which can demonstrate the comprehension ability of a model towards user behavior sequences. 
Three different models, including SIM (full-shot), Vicuna-13B (zero-shot) and ReLLa (few-shot), are evaluated with different $K$s. 
On BookCrossing dataset, $K$ ranges in $\{10, 20, 30, 40, 50, 60\}$, while on MovieLens-1M and MovieLens-25M datasets, $K$ ranges in $\{5, 10, 15, 20, 25, 30\}$. 
The numbers of shots are set to $256$, $8192$, $8192$ for BookCrossing, MovieLens-1M, and MovieLens-25M, respectively (\ie, <1\% few-shot setting).
The results are shown in Figure~\ref{fig:vary K}, from which we obtain the following observations:

\begin{figure*}[t]
\centering
\includegraphics[width=0.965\textwidth]{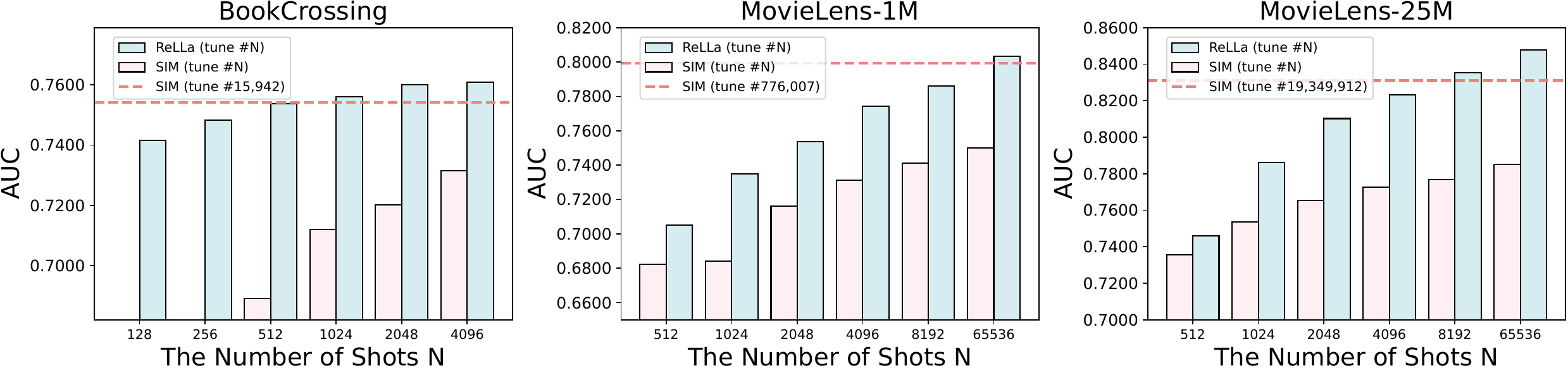}
\vspace{-5pt}
  \caption{The AUC performance of ReLLa and SIM w.r.t. different numbers of shots $N$ on three datasets, where ``tune \#$N$'' indicates that we train the model with $N$ training samples. 
  The dashed line denotes the AUC performance of SIM (full-shot) that is trained with the whole training set. 
  Notably, for $N=128$ and $N=256$ on BookCrossing dataset, few-shot SIM fails to accomplish the CTR prediction task, where the AUC is merely around 0.5, and is therefore omitted in the figure.
  }
  \vspace{-5pt}
  \label{fig:vary shot}
\end{figure*}

\begin{itemize}[leftmargin=10pt]
    \item As a traditional CTR prediction model, SIM (full-shot)~\cite{SIM} enjoys steady performance improvement as the length $K$ grows. 
    This is consistent with our common understanding, where longer user behavior sequences can introduce more useful information to better accomplish the recommendation tasks. 
    \item However, the performance of Vicuna-13B (zero-shot) only arrives at the peak with $K=30/15/15$ on BookCrossing/MovieLens-1M/MovieLens-25M datasets, and then starts to decrease with further longer sequence. 
    It is worth noting that the number of involved tokens (\ie, around 500/700/700 for three datasets respectively) is actually far from reaching the context limitation of Vicuna-13B (\ie, 2048 tokens). 
    This indicates that it is non-trivial for LLMs to comprehend the textual context of long behavior sequences for recommendation, where a certain amount of in-domain knowledge is required.
    \item ReLLa mitigates the incomprehension problem of LLMs on long user behavior sequences for recommendation. 
    Compared with Vicuna-13B (zero-shot), whose performance drops when $K>30$ on BookCrossing and $K>15$ on MovieLens-1M and MovieLens-25M, there are no performance turning points for ReLLa. 
    Similar to SIM, the AUC performance of ReLLa achieves continuous improvement as $K$ grows, validating the comprehension ability of ReLLa for the textual contexts with longer behavior sequences.
\end{itemize}

\subsection{Data Efficiency (RQ3)}
\label{sec:study on shot N}

Focusing on few-shot settings, we investigate the data efficiency property by varying the number of shots $N$. 
In Figure~\ref{fig:vary shot}, we report the AUC performance of ReLLa and SIM (the best full-shot baseline) with different $N$s. 
For BookCrossing dataset, $N$ ranges in $\{128, 256, 512, 1024, 2048, 4096\}$. 
For MovieLens-1M and MovieLens-25M datasets, $N$ ranges in $\{512, 1024, 2048, 4096, 8192, 65536\}$. 
The length of user behavior sequence $K$ is set to $K=60/30/30$ for BookCrossing/MovieLens-1M/MovieLens-25M datasets, respectively.

As depicted in Figure~\ref{fig:vary shot}, both ReLLa and SIM attain performance enhancement as the number of shots $N$ gradually grows. 
However, with the same number of shots $N$, ReLLa can outperform SIM significantly and consistently by a large margin. 
Moreover, when $N$ is extremely small (\eg, 128 and 256) on BookCrossing dataset, SIM even fails to accomplish the CTR prediction task where AUC is merely around 0.5. 
With limited number of training samples, ReLLa shows remarkable data efficiency property and display considerable few-shot inference ability due to the intrinsic logical reasoning abilities and possession of open-world knowledge of LLMs.

\begin{table*}
\vspace{-7pt}
\caption{ The performance of different variants of ReLLa. We remove different components of ReLLa to evaluate the contribution of each part to the model. 
The best result is given in bold, and the second-best value is underlined. 
}
\vspace{-10pt}
\label{tab:ablation}
\resizebox{0.88\textwidth}{!}{
\renewcommand\arraystretch{1.1}
\begin{tabular}{c|ccc|ccc|ccc}
\toprule
\hline

\multicolumn{1}{c|}{\multirow{2}{*}{Model Variant}} & \multicolumn{3}{c|}{BookCrossing} & \multicolumn{3}{c|}{MovieLens-1M} & \multicolumn{3}{c}{MovieLens-25M} \\ 
\multicolumn{1}{c|}{} & AUC  & Log Loss & ACC & AUC  & Log Loss & ACC & AUC  & Log Loss & ACC \\ 
   \hline 
ReLLa (Ours) & \textbf{0.7482} & \underline{0.6265} & \textbf{0.6800} & \textbf{0.7927} & \textbf{0.5475} & \textbf{0.7196} & \textbf{0.8352} & \textbf{0.4693} & \textbf{0.7779} \\ 
ReLLa (w/o Mixture) & 0.7399 & \textbf{0.6002} & 0.6715 & 0.7849 & \underline{0.5693} & 0.6985 & 0.8192 & 0.4904 & \underline{0.7715} \\ 
ReLLa (w/o Retrieval) & 0.7167 & 0.9293 & 0.4898 & 0.7718 & 0.5795 & \underline{0.7039} & 0.8174 & \underline{0.4892} & 0.7685 \\ 
ReLLa ($\frac{1}{2}N$-shot) & \underline{0.7415} & 0.6268 & 0.6462 & \underline{0.7862} & 0.5781 & 0.6964 & \underline{0.8231} & 0.5157 & 0.7672 \\ 
ReLLa (w/o IT) & 0.7253 & 0.9277 & 0.5750 & 0.7013 & 0.6250 & 0.6507 & 0.7324 & 0.5858 & 0.7027 \\ 
ReLLa (w/o IT \& Retrieval) & 0.7176 & 0.9507 & 0.5649 & 0.6993 & 0.6291 & 0.6493 & 0.7503 & 0.6308 & 0.6427 \\ 
  
   \hline  
   \bottomrule          
\end{tabular}
\vspace{-10pt}
}
\end{table*}

\subsection{Ablation Study (RQ4)}
\label{sec:ablation}

To analyze the efficacy of each component in our proposed ReLLa framework, we design the following model variants of ReLLa. 
We set $N=256/8192/8192$ (<1\% setting) and $K=60/30/30$ for BookCross-ing/MovieLens-1M/MovieLens-25M datasets, respectively.
\begin{itemize}[leftmargin=10pt]
    \item \textbf{ReLLa (Ours)} is the complete version of our proposed method. The training data consists of both original and retrieval-enhanced samples, resulting in a mixed training dataset of $2N$ samples. 
    The testing set only contains pure retrieval-enhanced samples.
    \item \textbf{ReLLa (w/o Mixture)}. 
    We only maintain the retrieval-enhanced data instances to construct the training dataset of $N$ samples.
    The testing data is still all retrieval-enhanced samples.
    \item \textbf{ReLLa (w/o Retrieval)}. We remove the semantic user behavior retrieval for both training and testing samples. That is, training and testing data are all original samples without retrieval enhancements. The training set contains $N$ training samples. 
    This variant indicates the vanilla instruction tuning version over Vicuna-13B, which is similar to TALLRec~\cite{bao2023tallrec}.
    \item \textbf{ReLLa ($\frac{1}{2}N$-shot)}. We halve the number of shots $N$ to $\frac{1}{2}N$, \ie, from 256 to 128 on BookCrossing and from 8192 to 4096 on MovieLens-1M and MovieLens-25M. 
    Therefore, the constructed mixed training set contains $N$ training samples. 
    This variant is intended to decouple and investigate the factors of doubled training samples and pattern enrichment.
    \item \textbf{ReLLa (w/o IT)}. We remove the instruction tuning, while preserving the retrieval-enhanced samples for testing data. This variant indicates the zero-shot version of our proposed ReLLa.
    \item \textbf{ReLLa (w/o IT \& Retrieval)}. We remove both the instruction tuning and retrieval operation. Therefore, the testing data only contains original data samples. This variant indicates the zero-shot version of vanilla Vicuna-13B.
\end{itemize}

The performance of these variants are presented in Table~\ref{tab:ablation}, from which we can draw the following observations:
\begin{itemize}[leftmargin=10pt]
    \item For ReLLa (w/o Mixture) and ReLLa (w/o Retrieval), their training and testing data comprise exactly the same type of samples, \ie, either pure original samples or retrieval-enhanced samples respectively, which indicates that there is no data inconsistency between the training and testing phases. 
    Nevertheless, both of them significantly underperform our proposed ReLLa by at least 1.12\%, 0.99\% and 1.95\% on BookCrossing, MovieLens-1M and MovieLens-25M in AUC respectively. 
    This highlights the importance of the data mixture strategy, the benefits of which can be broken down into two prominent factors: doubled training samples and pattern enrichment. 
    Doubled training samples lead to a more thorough training process, while pattern enrichment can prevent the model from overfitting and therefore increase the model robustness. 
    \item We introduce the variant ReLLa ($\frac{1}{2}N$-shot) to further decouple and analyze the two factors mentioned above, \ie, doubled training samples and pattern enrichment. 
    Its total number of training samples is the same as those of ReLLa (w/o Mixture) and ReLLa (w/o Retrieval), except that ReLLa ($\frac{1}{2}N$-shot) loses the sight of half truly training instances. 
    In this case, ReLLa ($\frac{1}{2}N$-shot) still outperforms ReLLa (w/o Mixture) and ReLLa (w/o Retrieval) with 0.21\%, 0.16\% and 0.48\% relative AUC improvement, and achieves comparable or better performance in Log Loss and ACC. 
    This indicates that pattern enrichment as regularization plays a more vital role that contributes to the performance improvement.
    \item 
    Finally, comparing ReLLa (w/o IT) and ReLLa (w/o IT \& Retrieval), which fall back into zero-shot settings, we can observe that ReLLa (w/o IT) generally achieves significant improvements over ReLLa (w/o IT \& Retrieval), except for the AUC metric on MovieLens-25M.
    This demonstrates that semantic user behavior retrieval (SUBR) improves the quality of data samples and makes the filtered behavior sequence more friendly for LLM to extract useful knowledge.
\end{itemize}


\begin{figure}[t]
\centering
\vspace{-5pt}
\includegraphics[width=0.46\textwidth]{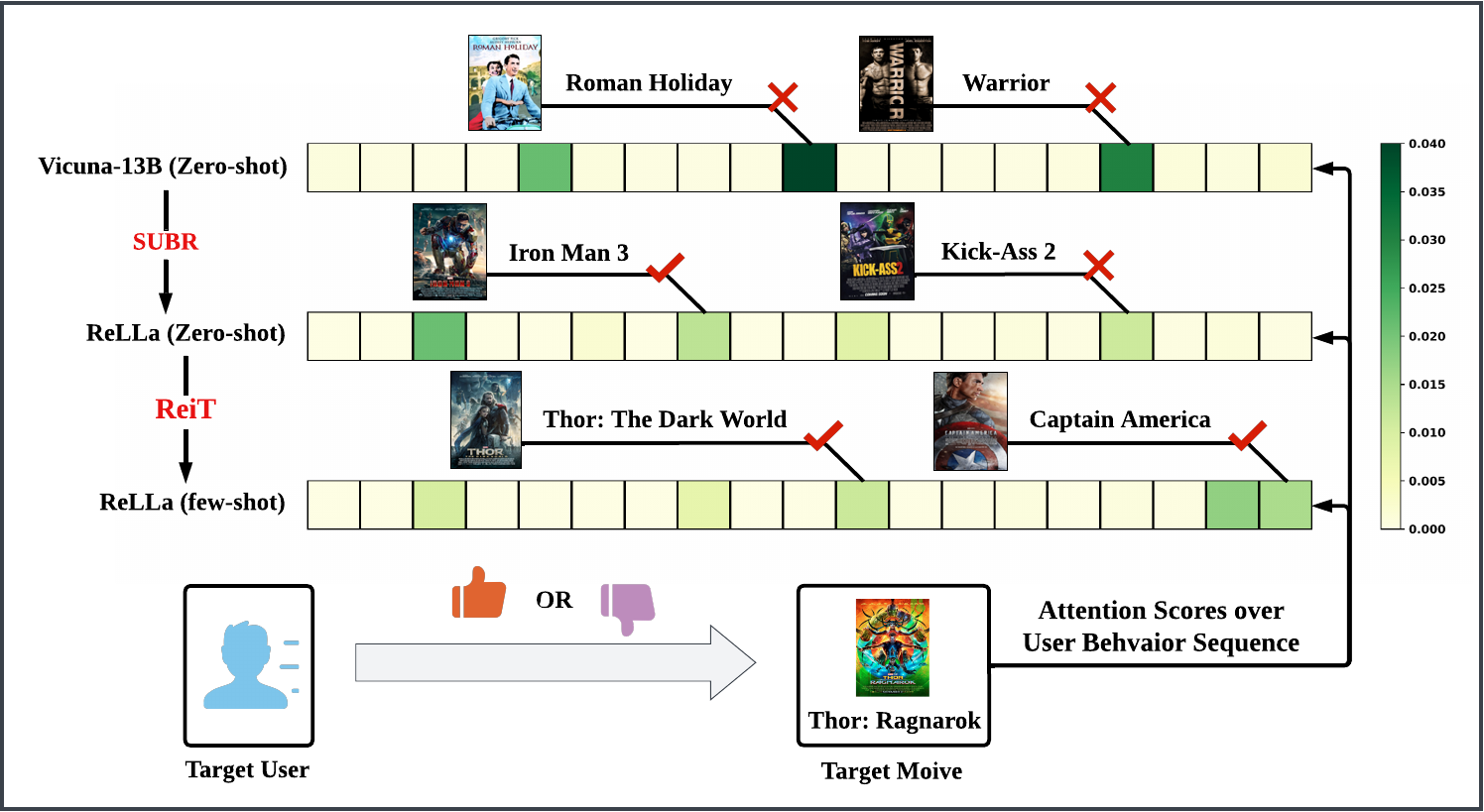}
\vspace{-5pt}
  \caption{
  The case study of ReLLa on MovieLens-25M dataset. We visualize the attention scores over the historical items (\ie, the rectangles) in user behavior sequence at the last hidden layer of LLM. 
  The deeper green a rectangle possesses, the large attention score the corresponding historical item attains, thus contributing more to the final CTR estimation.
  }
\vspace{-10pt}
  \label{fig:case study}
\end{figure}

\subsection{Case Study (RQ5)}
\label{sec: case study}

In this section, we conduct case study to further analyze how can ReLLa help LLM better understand the long user behavior sequence.
As shown in Figure~\ref{fig:case study}, we select a testing sample from MovieLens-25M dataset, and visualize the attention scores of target item over the user behavior sequence at the last hidden layer of three different models (\ie, Vicuna-13B, ReLLa (zero-shot), and ReLLa (few-shot)). 
The attention score for each historical item is computed by summing up the attention scores of every word token for the textual input of the corresponding item.
In Figure~\ref{fig:case study}, each historical item is represented as a rectangle with color ranging from yellow to green.
The deeper green a rectangle possesses, the large attention score the corresponding historical item attains, thus contributing more to the final CTR estimation.

For Vicuna-13B (zero-shot), the largest attentions fall on the movie \textit{Roman Holiday} and \emph{Warrior}, which have little relationship with the target movie \textit{Thor: Ragnarok}, and thus the model fails to correctly infer the user's preference towards the target item. 
Equipped with semantic user behavior retrieval (SUBR), we can reduce the noise of user behavior sequence and bring in more relevant items. 
As shown in Figure~\ref{fig:case study}, ReLLa (zero-shot) is able to put more attentions to superhero movies (\eg, \emph{Iron Man 3}) that are semantically similar to the target item. 
However there are still outliers for ReLLa (zero-shot), \eg, the movie \emph{Kick-Ass 2} is generally non-correlated to \textit{Thor: Ragnarok} produced the Marvel. 
Next, by further applying retrieval-enhanced instruction tuning (ReiT), we can observe that the large attention weights of ReLLa (few-shot) all fall onto relevant superhero movies that are also produced by the Marvel.
Therefore, we can conclude that our proposed SUBR and ReiT can help LLM to correctly grasp the correlation between the target item and historical items, thus better comprehending the user behavior sequence.

\section{Related Work}
\label{sec: related work}

\subsection{Traditional CTR Prediction}

CTR prediction serves as the key component for various online applications (\eg, recommender systems~\cite{xi2023bird}, advertising~\cite{ou2023deep}, and web search~\cite{lin2021graph,fu2023f,dai2021adversarial}).
It aims to accurately estimate the user's click probability towards a certain target item in a given context~\cite{zhang2021deep}.
Traditional CTR prediction models can be mainly classified into two categories: (1) feature interaction based models, and (2) sequential recommendation models.

The feature interaction based models generally derive from POLY2~\cite{POLY2} and FM~\cite{FM}. 
Their core idea is to capture the second- or high-order feature interaction patterns across multiple feature fields with different operators (\eg, product~\cite{PNN,DCNv2,DeepFM,EDCN}, convolution~\cite{CFM,FGCNN}, and attention~\cite{AutoInt,AFM}). 
For examples, DCN~\cite{DCNv1}, xDeepFM~\cite{xDeepFM}, and DCNv2~\cite{DCNv2} apply product-based feature crossing operation at each layer for explicit high-order feature interaction modeling.
AutoInt~\cite{AutoInt} and InterHAt~\cite{InterHAt} adopt the attention mechanism for feature interactions, which provides additional explainable prediction via attention weights.

The sequential recommendation model~\cite{zhou2019deep,pi2019practice,zhou2018deep} focuses on user behavior modeling and seeks to dynamically capture users' interests towards a target item according to the given behavior history.
They leverage different architectures (\eg, RNN~\cite{hidasi2017recurrent,GRU4Rec}, CNN~\cite{Caser}, attention~\cite{zhou2019deep,zhou2018deep}, memory bank~\cite{pi2019practice,ren2019lifelong}) to handle the user behavior sequence for user preference modeling.
For instances, GRU4Rec~\cite{GRU4Rec} adopt the gated recurrent unit (GRU)~\cite{chung2014empirical} to encode the user's sequential behaviors. 
Caser~\cite{Caser} introduces the convolution neural network (CNN) to model the union-level patterns among user behavior sequences.

\subsection{Language Models for Recommendation}

As suggested in previous work~\cite{lin2023can}, the adaption of language models to the field of recommender systems can be generally categorized according to the roles they serve in the recommendation pipeline, \ie, feature engineering~\cite{liu2023first,borisov2022language,li2023taggpt,mysore2023large,carranza2023privacy,christakopoulou2023large}, feature encoder~\cite{muhamed2021ctr,hou2022towards,yu2021tiny,wang2022transrec,hou2023learning,zhang2022twhin,fu2023exploring,yuan2023go,qiu2021u,li2023exploring}, scoring/ranking function~\cite{liu2022ptab,kang2023llms,zhang2021language,li2023pbnr,bao2023tallrec,li2023text,zhang2023prompt,mao2023unitrec,hua2023up5,geng2023vip5,hua2023index,zhang2023chatgpt,hou2023large,chen2023palr,petrov2023generative,wang2023zero}.

For feature engineering, large language models (LLMs) accept the raw data (\eg, user profiles and item descriptions) as input, and generate supplementary text-based attributes as data augmentations with delicately designed prompts and templates. 
For example, KAR~\cite{xi2023towards} utilizes the reasoning knowledge on user preferences and the factual knowledge on items by requesting LLMs with factorization prompting techniques. 
The obtained knowledge can serve as augmented features and promote the recommendation performance in a model-agnostic manner. 
GENRE~\cite{liu2023first} employs LLMs to obtain news summarization, synthetic news pieces, and user profiles. 

For feature encoder, LLMs are adopted as auxiliary textual feature encoders to (1) enrich the user/item representations with semantic information, and (2) enable cross-domain recommendation with the natural language interface. 
For instance, U-BERT~\cite{qiu2021u} enhances the user representation by encoding review texts into dense vectors via BERT. UniSRec~\cite{hou2022towards} and VQ-Rec~\cite{hou2023learning} apply a fixed BERT as the encoder for item descriptive texts, in order to achieve unified cross-domain sequential recommendation.

For scoring/ranking function, researchers explore the potential of LLMs to directly serve as the core scoring or ranking module for recommendation, instead of an assistant role for conventional recommendation models (\eg, feature engineering or feature encoder). 
In this case, LLMs are employed to accomplish either the item scoring task~\cite{liu2022ptab,kang2023llms,zhang2021language,li2023pbnr,bao2023tallrec,li2023text,zhang2023prompt,mao2023unitrec}, or item generation task~\cite{hua2023up5,geng2023vip5,hua2023index,zhang2023chatgpt,hou2023large,chen2023palr,petrov2023generative,wang2023zero}. 
Also, various works~\cite{geng2022recommendation,cui2022m6,zhang2023recommendation,liu2023chatgpt,sun2023chatgpt,dai2023uncovering} attempt to utilize the multi-task capacity of LLMs, and instruct LLMs to solve the multiple tasks (\eg, both scoring and generation) through a unified language interface.

In this paper, we mainly focus on the utilization of LLMs as the scoring/ranking functions, where the pointwise scoring task is adopted for CTR prediction. 
To the best of our knowledge, we are the first to identify and well formulate the incomprehension problem of LLMs on lifelong user behavior sequences when adopting LLMs for scoring and ranking tasks. 
A novel ReLLa framework is proposed to mitigate such an issue by introducing the retrieval techniques to promote comprehension ability of LLMs and thus enhance their recommendation performance.

\section{Conclusion}

In this paper, we focus on adapting and empowering LLMs as the scoring/ranking function for recommendation tasks.
We first identify and formulate the incomprehension problem of LLMs on lifelong sequential behaviors, \ie, LLMs fail to extract useful information from a textual context of long user behavior sequence, even if the length of context is far from reaching the context limitation of LLMs. 
Hence, we propose a novel ReLLa framework, where semantic user behavior retrieval (SUBR) and retrieval-enhanced instruction tuning (ReiT) are designed to address such an issue and therefore promote the recommendation performance. 
Extensive experiments validate the effectiveness of our proposed ReLLa compared with existing baselines. Specifically, leveraging only less than 10\% training samples, \textit{few-shot} ReLLa can outperform all the \emph{full-shot} traditional CTR models that are trained on the entire training set. 
This demonstrate the superior data efficiency of ReLLa, as well as its comprehension ability towards long user behavior sequences.

\begin{acks}
The Shanghai Jiao Tong University team is partially supported by National Key R\&D Program of China (2022ZD0114804), Shanghai Municipal Science and Technology Major Project (2021SHZDZX0102) and National Natural Science Foundation of China (62177033, 62322603). 
The work is sponsored by Huawei Innovation Research Program.
We thank MindSpore~\cite{mindspore} for the partial support of this work, which is a new deep learning computing framework.
\end{acks}

\bibliographystyle{ACM-Reference-Format}
\bibliography{acmart}

\appendix

\section{Prompt Illustration}
\label{app:prompt}

We demonstrate several examples to illustrate the hard prompt templates used for ReLLa on all three datasets.  

\begin{figure}[t]
\centering
\vspace{-7pt}
\includegraphics[width=0.48\textwidth]{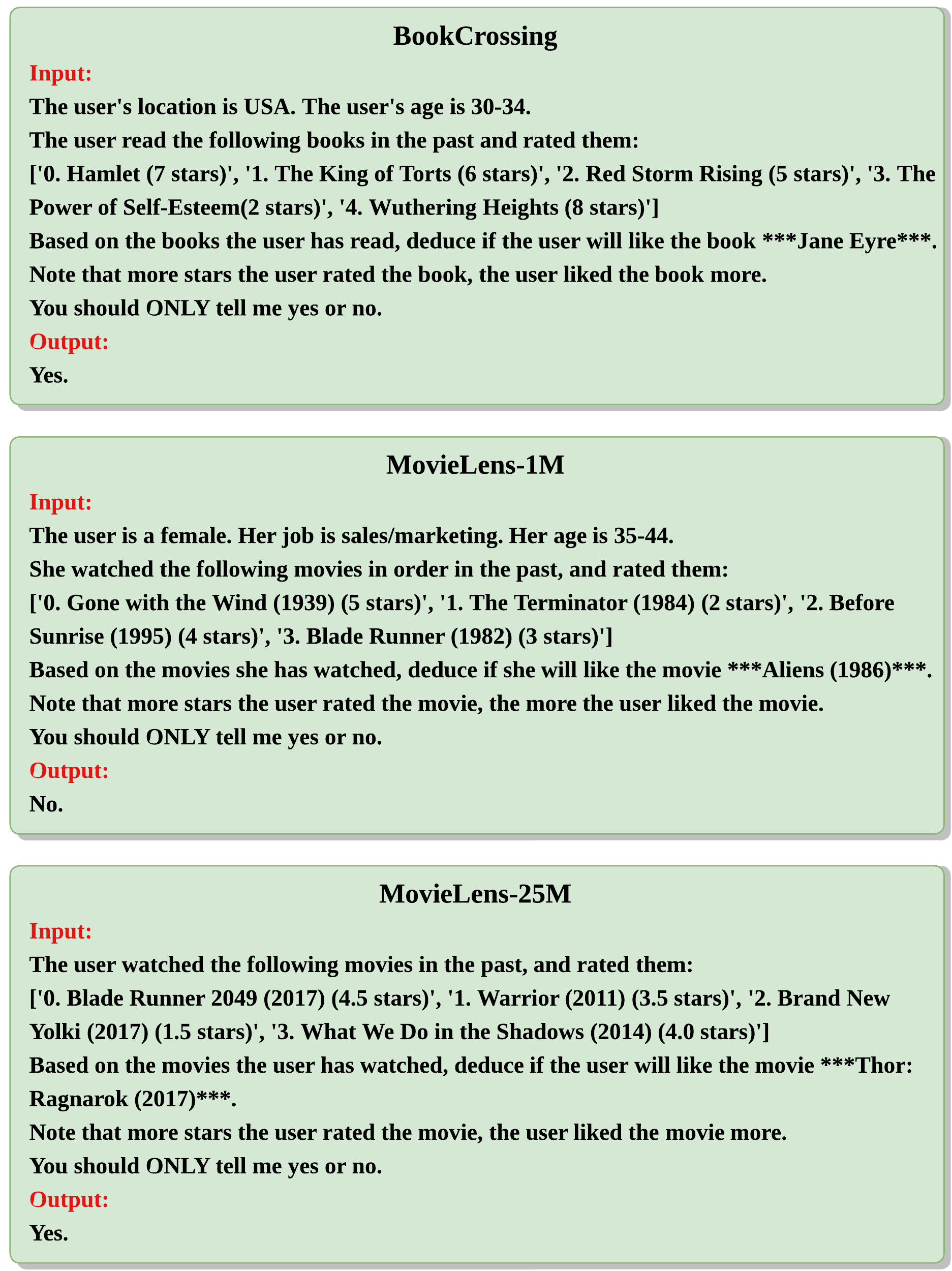}
  \caption{Examples of hard prompt templates for three datasets \emph{without} SUBR. The user behavior sequence is constructed by the most recent $K$ items.
  }
  \vspace{-10pt}
  \label{fig: prompt simple}
\end{figure}

\begin{figure}[t]
\centering
\vspace{-7pt}
\includegraphics[width=0.48\textwidth]{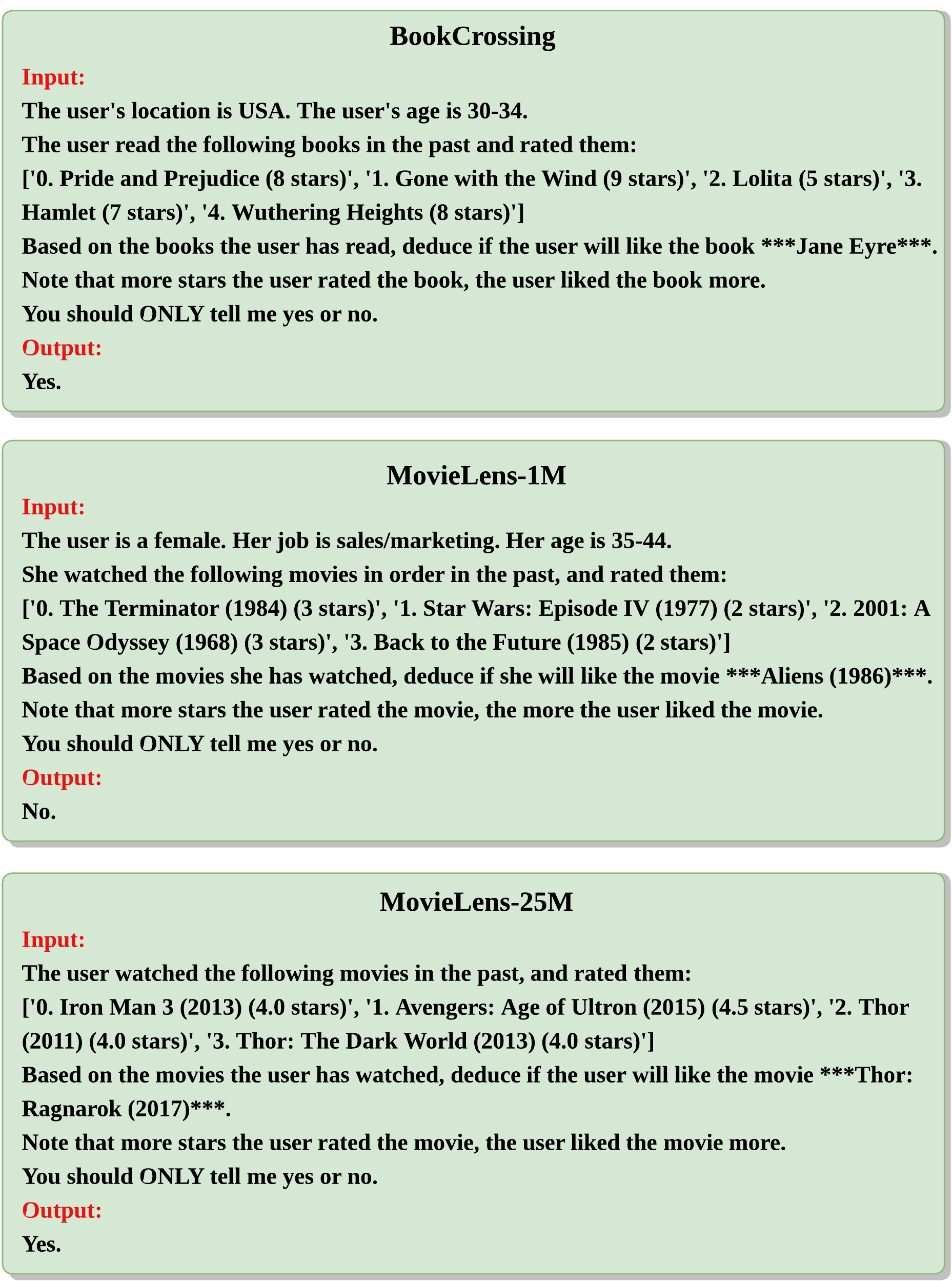}
  \caption{Examples of hard prompt templates for three datasets \emph{with} SUBR. The user behavior sequence is constructed by the most relevant $K$ items.
  }
  \label{fig: prompt ret}
\end{figure}

Figure~\ref{fig: prompt simple} shows the textual input-output pairs without semantic user behavior retrieval (SUBR), where the user behavior sequence is truncated to most recent $K$ (\eg, $K=4$ in the figure). 
As is shown in Figure~\ref{fig: prompt ret}, after applying SUBR, the user behavior sequence will be replaced by most relevant $K$ historical items towards the target item.
For example, for MovieLens-25M dataset, historical behaviors retrieved by SUBR are all related to superheros or Marvel, which is highly correlated to the target movie ``Thor: Ragnaro''. 
Note that the user behavior sequence generated by SUBR keeps the chronological  order in the original lifelong user sequence.  

Figure~\ref{fig:item description} demonstrates how we design prompts for item descriptions on the three datasets, which will be encoded by LLM for semantic user behavior retrieval (SUBR). 

\begin{figure}[t]
\centering
\includegraphics[width=0.48\textwidth]{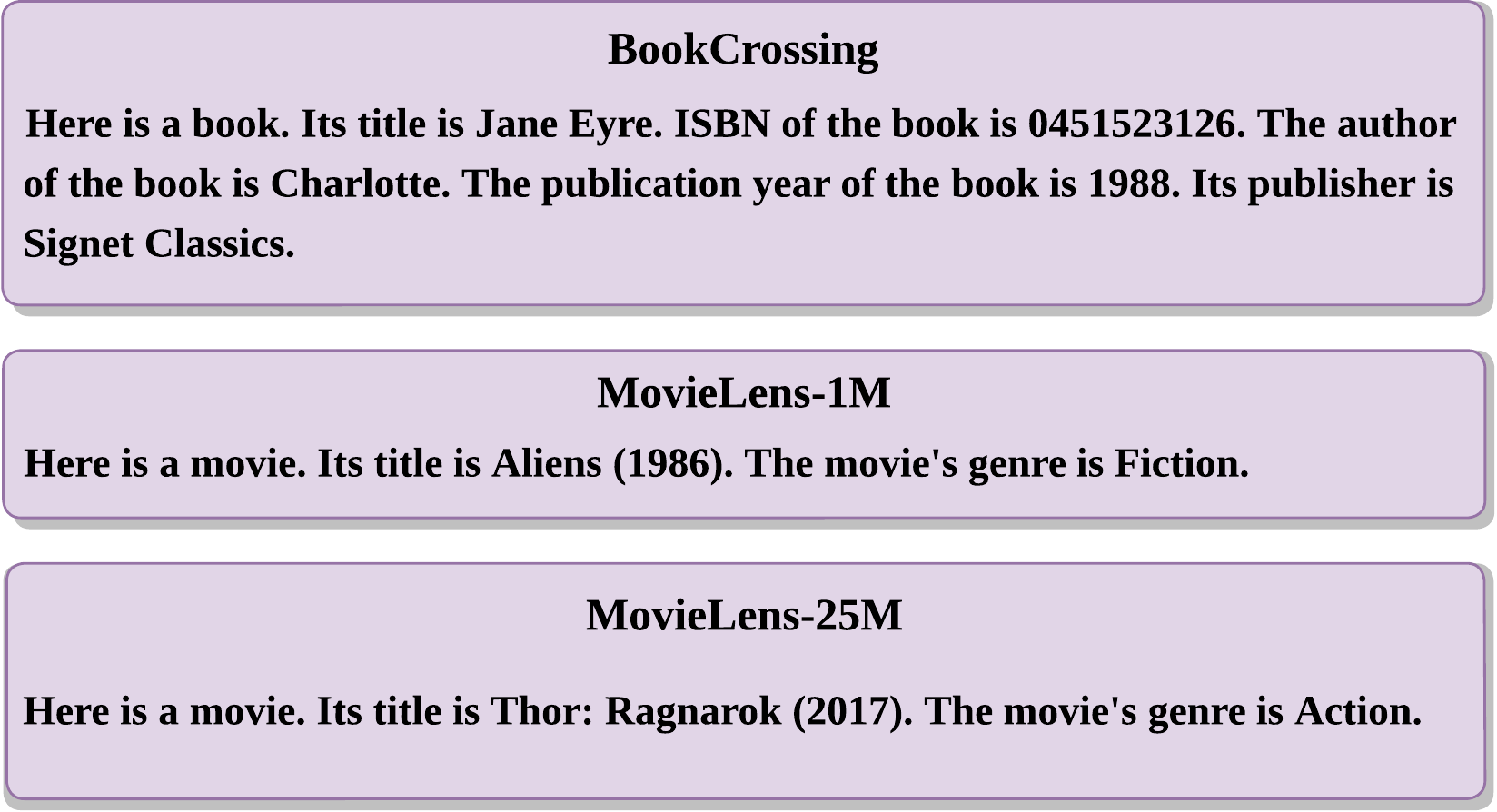}
  \caption{Examples of hard prompt templates of item descriptions for three datasets. The textual description is used to obtain the semantic item embedding from LLM, which will then be leveraged by SUBR.
  }
  \label{fig: full item description}
\end{figure}

\section{Data Preprocessing}
\label{app:dataset}
Our experiments are conducted on three real-world public datasets (\ie, BookCrossing, MovieLens-1M and MovieLens-25M), and the statistics of the processed datasets are show in Table~\ref{tab:datasets}. MovieLens-1M and MovieLens-25M datasets are split into training and testing sets with ratio of 8:1 according to the global timestamp~\cite{qin2021retrieval}. 
Since BookCrossing dataset has no timestamps, following previous work~\cite{bao2023tallrec}, we divide it into training and testing sets with ratio of 9:1 by random split of users.  Data samples with user behavior sequence length less than 5 are filtered on all three datasets. We describe more preprocessing details as follows:

\begin{itemize}[leftmargin=10pt]
    \item \textbf{BookCrossing} possesses user-book integer ratings ranging from 0 to 10. We consider samples with rating above 5 as positive, and the rest as negative. 
    \item \textbf{MovieLens-1M} contains user-movie integer ratings ranging from 0 to 5. Samples with ratings of 4 and 5 are labeled as positive and the rest as negative.~\cite{zhou2018deep,xi2023towards}
    \item \textbf{MovieLens-25M} has a scoring range from 0 to 5, with increments of 0.5. We label samples with ratings above 3.0 as positive, and the rest as negative.
\end{itemize}

Under the few-shot setting with a particular number of shot $N$, we uniformly sample $N$ data instances from the training set, which is then fixed ReLLa during few-shot tuning.
Note that the sampled data instances with a smaller $N$ are all included in the sampled few-shot training sets with a larger $N$.

\section{Baseline Implementation}
\label{app:baseline}

In this section, we describe the hyperparameter configuration for the baseline models from two different categories: (1) traditional CTR models, and (2) LM-based models.

\subsection{Traditional CTR Models}
We choose the embedding size from \{8, 16, 32\} on BookCrossing dataset and \{32, 64\} on MovieLens-1M and MovieLens-25M datasets. The dropout rate is selected from \{0.0, 0.1, 0.2\}. The activation function is fixed to ReLU. The learning rate is set to $1\times 10^{-3}$ and AdamW~\cite{adamw} optimizer is used. On BookCrossing, the batch size is selected from \{32, 64\}. On MovieLens-1M and MovieLens-25M, the batch size is selected from \{256, 512\}. More model-specific hyperparameter settings are shown as follows:

\begin{itemize}[leftmargin=10pt]
    \item \textbf{DeepFM}~\cite{DeepFM}. On BookCrossing, the size of DNN layer is selected from \{32, 64, 128\}. The number of DNN layers is selected from \{1, 2, 3\}. On MovieLens-1M and MovieLens-25M, we choose the size of DNN layer from \{128, 256\} and the number of DNN layers from \{3, 6, 9, 12\}.
    \item \textbf{AutoInt}~\cite{AutoInt}. On BookCrossing, the number of attention layers is selected from \{1, 2\} and the attention size is set to 32. On MovieLens-1M and MovieLens-25M, the attention layers is selected from \{3, 6, 9, 12\} and the attention size is selected from \{64, 128, 256\}. The number of attention heads are all set to 1.
    \item \textbf{DCNv2}~\cite{DCNv2}. On BookCrossing, the size of DNN layer is selected from \{32, 64, 128\}. The number of DNN layers and cross layers are selected from \{1, 2, 3\}. On MovieLens-1M and MovieLens-25M, we choose the size of DNN layers from \{128, 256\} and the number of DNN layers and cross layers are from \{3, 6, 9, 12\}.
    \item \textbf{GRU4Rec}~\cite{GRU4Rec}. The number of GRU layers is selected from \{1, 2, 3\}. On BookCrossing, the GRU hidden size and DNN hidden size is selected from \{32, 64\}. On MovieLens-1M and MovieLens-25M, the GRU hidden size and DNN hidden size is selected from \{64, 128, 256\}.
    \item \textbf{Caser}~\cite{Caser}. The number of vertical convolution kernels is selected from \{2, 4, 8\}. The number of horizontal convolution kernels is selected from \{4, 8, 16\}. The number of DNN layers is selected from \{1,2,3\}. The DNN hidden size is selected from \{32, 64\} on BookCrossing and \{64, 128, 256\} on MovieLens-1M and MovieLens-25M.
    \item \textbf{SASRec}~\cite{SASRec}. The number of attention heads is selected from \{1, 2, 4\}. The number of attention layers is selected from \{1, 2, 3\}. The attention size is selected from \{32, 64, 128\} on BookCrossing and \{64, 128, 256\}. The number of DNN layers is selected from \{1,2,3\}. The DNN hidden size is selected from \{32, 64\} on BookCrossing and \{64, 128, 256\} on MovieLens-1M and MovieLens-25M.
    \item \textbf{DIN}~\cite{zhou2018deep}. The number of DIN attention layers and DNN layers are selected from \{1, 2, 3\}. The DNN hidden size is selected from \{32, 64\} on BookCrossing and \{64, 128, 256\} on MovieLens-1M and MovieLens-25M.
    \item \textbf{SIM}~\cite{SIM}. The number of attention layers and DNN layers are selected from \{1, 2, 3\}. The DNN hidden size is selected from \{32, 64\} on BookCrossing and \{64, 128, 256\} on MovieLens-1M and MovieLens-25M.
\end{itemize}

\subsection{LM-based Models}
The structure of the pretrained language models is kept unchanged. And AdamW~\cite{adamw} optimizer is used for all the baselines. The detailed training settings are as follows:

\begin{itemize}[leftmargin=10pt]
    \item \textbf{CTR-BERT}~\cite{CTRBERT}. We maintain a two-tower model structure based on the BERT~\cite{devlin2018bert} model to encode the user and item information respectively. The total number of tuning epochs is set to 10. The batch size is set to 1024. The learning rate is set to $5\times 10^{-5}$ with linear decay. The warmup ratio is 0.05.
    \item \textbf{P5}~\cite{P5} is a unified sequence-to-sequence framework with T5~\cite{T5} as the backbone pretrained language model for multiple recommendation tasks. In this paper, we leverage P5 for a single task only (\ie, CTR prediction). The total number of epochs is set to 10 with batch size of 32. The learning rate is selected from
$\{5 \times 10^{-4}, 1 \times 10^{-3}\}$ with linear decay. The warmup ratio is 0.05. Following P5’s official implementation, we also perform gradient clip with threshold equal to 1.0.
    \item \textbf{PTab}~\cite{liu2022ptab} adopts the common pretrain-finetune scheme based
on the BERT~\cite{devlin2018bert} model. PTab first further pretrains the BERT model with the classical masked language modeling objective
based on the textualized CTR data, and then finetunes BERT
for downstream CTR prediction as a text classification problem.
Following the original paper, we pretrain BERT for 10 epochs with batch size equal to 1024. The learning rate for pretraining
is set to $5\times 10^{-5}$ with linear decay. The warmup ratio is 0.05. As
for finetuning, the total number of tuning epoch is set to 10 with
batch size of 1024. The learning rate for finetuning is initialized
at $5\times 10^{-5}$ with linear decay. The warmup ratio is 0.01.

\end{itemize}


\section{Additional Experiments}
\label{app:exp}

In this section, we further provide additional experiments to verify the following core points:
\begin{itemize}[leftmargin=10pt]
    \item The \textit{universality} of the lifelong sequential behavior incomprehension problem and the \textit{generalization} of our proposed ReLLa.
    \item Analysis about the model parameter and inference time.
    \item Ablation on PCA dimensionality and distance metrics for SUBR.
    \item Analysis and discussion about the potential reason for the incomprehension problem.
\end{itemize}

\subsection{Universality \& Generalization}

We validate the universality of the lifelong sequential behavior incomprehension problem and he generalization of our proposed ReLLa, by incorporating different backbone LLMs of different architectures and sizes including Falcon-7B~\cite{almazrouei2023falcon}\footnote{\url{https://huggingface.co/tiiuae/falcon-7b-instruct}}, Mistral-7B~\cite{jiang2023mistral}\footnote{\url{https://huggingface.co/mistralai/Mistral-7B-Instruct-v0.1}}, Vicuna-7B~\cite{vicuna2023}\footnote{\url{https://huggingface.co/lmsys/vicuna-7b-v1.3}}, Vicuna-13B~\cite{vicuna2023}\footnote{\url{https://huggingface.co/lmsys/vicuna-13b-v1.3}}, LLaMA-2-70B-Chat~\cite{touvron2023llama}\footnote{\url{https://huggingface.co/meta-llama/Llama-2-70b-chat-hf}}.

\subsubsection{Universality of the Incomprehension Problem}

We first analyze the universality of lifelong sequential behavior incomprehension problem for different LLMs on MovieLens-1M dataset. 
We report the zero-shot AUC performance of different LLMs w.r.t. different length $K$ of user behavior sequence ranging from \{5, 10, 15, 20, 25, 30\}. 
It is worth noting that we downsample the test set to 10,000 for LLaMA2-70B-chat due to the time consumption, while the other four LLMs are still tested on the whole test set. 
The results are given in Table~\ref{tab:universality}. 
We can observe that the length of user sequence at which a peaking performance is reached is small and far from reaching the context limit for all the five LLMs. Therefore, we validate the universality of lifelong sequential behavior incomprehension problem when adapting LLMs to recommendation domains. Besides, there exists performance difference among different LLMs, which may be related to the inherent instruction-following capabilities of LLMs themselves.

\begin{table}[h]
    \caption{Zero-shot AUC performance w.r.t. different sequence length $K$ for different LLMs on MovieLens-1M dataset. The peaking performance for each LLM is given in bold.
    }
    \vspace{-10pt}
    \label{tab:universality}
    \resizebox{0.48\textwidth}{!}{
    \renewcommand\arraystretch{1.1}
    \begin{tabular}{c|ccccccc}
    \toprule
    \hline
    \multicolumn{1}{c|}{\multirow{2}{*}{LLM}} & \multicolumn{6}{c}{MovieLens-1M} \\ 
    \multicolumn{1}{c|}{} & K=5 & K=10 & K=15 & K=20 & K=25 & K=30 \\ 
   \hline 
   Falcon-7B & \textbf{0.5906} & 0.5741 & 0.5583 & 0.5420 & 0.5468 & 0.5452 \\ 
   Mistral-7B & 0.6566 & 0.6568 & \textbf{0.6670} & 0.6623 & 0.6612 & 0.6610 \\
   Vicuna-7B & 0.6630 & 0.6586 & \textbf{0.6739} & 0.6527 & 0.6463 & 0.6412 \\
   Vicuna-13B & 0.6807 & 0.6932 & \textbf{0.6993} & 0.6918 & 0.6937 & 0.6908 \\
   LLaMA2-70B & 0.6259 & 0.6348 & \textbf{0.6421} & 0.6402 & 0.6339 & 0.6321 \\
   \hline  
   \bottomrule          
\end{tabular}
}
\end{table}

\subsubsection{Generalization of ReLLa}

We further investigate the generalization of our proposed ReLLa in terms of different backbone LLMs (\ie, model compatibility). 
We apply semantic user behavior retrieval (SUBR) and retrieval-enhanced instruction tuning (ReiT) on four LLMs, excluding LLaMA-2-70B-Chat due to the computational resource constraint. 
The finetuning configuration is set as the same as our previous experiment on Vicuna-13B. We set the length of user sequence to 30. 
Few-shot settings <1\% and <10\% indicate 8,192 and 65,536 training samples, respectively. 
We also provide the performance of the best baseline model (i.e., SIM) for both full-shot and few-shot settings. 
We report the results in Table~\ref{tab:generalization}, from which we have the following observations and discussions:
\begin{itemize}[leftmargin=10pt]
    \item Compared with the original LLMs, ReLLa can improve the recommendation performance in both zero-shot and few-shot settings consistently and significantly.
    \item Mistral-7B, Vicuna-7B and Vicuna-13B with ReiT (<10\%) settings are able to significantly outperform full-shot SIM that is trained on the whole training set, which demonstrates the surprisingly high data efficiency property of ReLLa.
    \item Although Falcon-7B with ReiT (<10\%) setting obtains much better performance than SIM under the same few-shot setting (<10\%), it fails to defeat full-shot SIM as other three LLMs do. We think the main reasons are in two folds:
    \begin{enumerate}
        \item \textit{Model Capability}. The recommendation performance of LLM is highly correlated to its instruction-following capability. Maybe Falcon-7B itself is not ready to suit recommendation tasks.
        \item \textit{Tuning Strategy}. We simply set the hyperparameter configuration and prompt template to be the same as Vicuna-13B, which might be suboptimal. We will proceed to optimize the performance to see whether few-shot Falcon-7B can defeat full-shot SIM or not.
    \end{enumerate}
    \item While the peaking performance arrives with $K$=5 or 15 for zero-shot LLMs, the four LLMs gain much better recommendation performance with larger $K$=30 when equipped with ReLLa. Hence, we validate the generalization of ReLLa to address the user sequence incomprehension problem and improve the recommendation performance.
\end{itemize}

\begin{table}[t]
    \caption{The model compatibility of ReLLa w.r.t. different backbone LLMs on MovieLens-1M dataset with $K$=30. We also give the performance of SIM, which is the best baseline among traditional recommendation models.
    }
    \vspace{-10pt}
    \label{tab:generalization}
    \resizebox{0.48\textwidth}{!}{
    \renewcommand\arraystretch{1.1}
    \begin{tabular}{c|c|ccc}
    \toprule
    \hline
    \multicolumn{2}{c|}{\multirow{2}{*}{Model}} & \multicolumn{3}{c}{MovieLens-1M} \\ 
    \multicolumn{2}{c|}{} & AUC & Log Loss & ACC \\ 
   \hline 
   \multicolumn{1}{c|}{\multirow{3}{*}{SIM}} & few-shot (<1\%) & 0.7352 & 0.6132 & 0.6743 \\ 
   \multicolumn{1}{c|}{} & few-shot (<10\%) & 0.7414 & 0.6129 & 0.6756 \\ 
   \multicolumn{1}{c|}{} & full-shot & \textbf{0.7992} & \textbf{0.5387} & \textbf{0.7268} \\ 
   \hline 
   \multicolumn{1}{c|}{\multirow{4}{*}{Falcon-7B}} & zero-shot & 0.5906 & 0.7674 & 0.5436 \\ 
   \multicolumn{1}{c|}{} & with SUBR & 0.5964 & 0.7709 & 0.5437 \\ 
   \multicolumn{1}{c|}{} & with ReiT (<1\%) & 0.7811 & 0.5589 & 0.7111 \\ 
   \multicolumn{1}{c|}{} & with ReiT (<10\%) & \textbf{0.7870} & \textbf{0.5658} & \textbf{0.7072} \\ 
   \hline 
   \multicolumn{1}{c|}{\multirow{4}{*}{Mistral-7B}} & zero-shot & 0.6670 & 0.7556 & 0.4793 \\ 
   \multicolumn{1}{c|}{} & with SUBR & 0.6881 & 0.7321 & 0.5119 \\ 
   \multicolumn{1}{c|}{} & with ReiT (<1\%) & 0.7905 & 0.5488 & 0.7210 \\ 
   \multicolumn{1}{c|}{} & with ReiT (<10\%) & \textbf{0.8005} & \textbf{0.5388} & \textbf{0.7275} \\ 
   \hline 
   \multicolumn{1}{c|}{\multirow{4}{*}{Vicuna-7B}} & zero-shot & 0.6739 & 0.9510 & 0.5644 \\ 
   \multicolumn{1}{c|}{} & with SUBR & 0.6704 & 0.7745 & 0.5655 \\ 
   \multicolumn{1}{c|}{} & with ReiT (<1\%) & 0.7918 & 0.5493 & 0.7196 \\ 
   \multicolumn{1}{c|}{} & with ReiT (<10\%) & \textbf{0.8016} & \textbf{0.5365} & \textbf{0.7274} \\ 
   \hline 
   \multicolumn{1}{c|}{\multirow{4}{*}{Vicuna-13B}} & zero-shot & 0.6993 & 0.6291 & 0.6493 \\ 
   \multicolumn{1}{c|}{} & with SUBR & 0.7013 & 0.6250 & 0.6507 \\ 
   \multicolumn{1}{c|}{} & with ReiT (<1\%) & 0.7927 & 0.5475 & 0.7196 \\ 
   \multicolumn{1}{c|}{} & with ReiT (<10\%) & \textbf{0.8033} & \textbf{0.5362} & \textbf{0.7280} \\ 
   \hline  
   \bottomrule          
\end{tabular}
}
\end{table}


\subsection{Model Parameter \& Inference Time}

We provide the complexity analysis on MovieLens-1M dataset by reporting the number of total parameters, the number of trainable parameters, and the averaged inference time per batch for both ReLLa and SIM (the best traditional recommendation baseline). 
We choose Vicuna-13B as the backbone LLM for ReLLa.
The evaluation batch size is set to 512 and 4 for SIM and ReLLa, respectively. 
The run-time experiment is exclusively conducted on the same server with one GeForce RTX 4090 GPU.

\begin{table}[t]
    \caption{Complexity analysis on MovieLens-1M dataset.
    }
    \vspace{-10pt}
    \label{tab:complexity}
    \resizebox{0.48\textwidth}{!}{
    \renewcommand\arraystretch{1.1}
    \begin{tabular}{cccc}
    \toprule
    \hline
    Model & \# Total Parameter & \# Trainable Parameter & Inference Time \\ 
    \hline
    SIM & 1.44M & 1.44M & 3.21ms \\
    ReLLa & 13B & 650M & 500ms \\
   \hline  
   \bottomrule          
\end{tabular}
    \vspace{-10pt}
}
\end{table}

We report the results in Table~\ref{tab:complexity}.
Although ReLLa achieves remarkable success on sequential recommendation in terms of both performance and sample efficiency, we have to admit that the inference speed is slower compared with traditional recommendation models. 
Hence, ReLLa is currently suitable for real-world applications with a high tolerance for latency, such as conversational search or conversational recommendation.
Note that this computational limitation inherently stems from the large-scale property of LLMs. 
It is not unique to ReLLa but rather a common issue that our research community of LLM for recommendation should make joint efforts to overcome.

\subsection{Ablation on PCA \& Distance Metric}

We conduct ablation study to investigate the impact of PCA dimensionality and distance metrics for SUBR, respectively.
We choose Vicuna-13B as the backbone LLM for ReLLa.

\subsubsection{Impact of PCA Dimensionality}

To offer a deeper insight into the impact of PCA dimensionality, we evaluate the performance of ReLLa w.r.t. different PCA dimensionalities on MovieLens-1M dataset under both zero-shot and few-shot (8192-shot, <1\%) settings. 
The results are reported in Table~\ref{tab:ablation pca}. 
We can observe that PCA dimensionality 512 generally achieves the best performance. The dimension reduction brought by PCA also means a kind of semantic information loss. 
Hence, the smaller the dimensionality, the worse ReLLa performs. 
While dimensionality larger than 512 might lead to heavy cost of storage and computing, we consider 512 as a reasonable choice of PCA dimensionality to balance the performance and storage/computing cost.

\begin{table}[t]
    \caption{Ablation study w.r.t different PCA dimensionalities for ReLLa on MovieLens-1M dataset under both zero-shot and few-shot (<1\%) settings.
    }
    \vspace{-5pt}
    \label{tab:ablation pca}
    \resizebox{0.45\textwidth}{!}{
    \renewcommand\arraystretch{1.1}
    \begin{tabular}{c|c|ccc}
    \toprule
    \hline
    \multicolumn{1}{c|}{\multirow{2}{*}{Setting}} & \multicolumn{1}{c|}{\multirow{2}{*}{PCA Dim.}} & \multicolumn{3}{c}{MovieLens-1M} \\ 
    \multicolumn{1}{c|}{} & \multicolumn{1}{c|}{} & AUC & Log Loss & ACC \\
    \hline
    \multicolumn{1}{c|}{\multirow{4}{*}{zero-shot}} & 512 & 0.7013 & \textbf{0.6250} & \textbf{0.6507} \\
    \multicolumn{1}{c|}{} & 256 & \textbf{0.7064} & 0.6377 & 0.6357 \\
    \multicolumn{1}{c|}{} & 128 & 0.7063 & 0.6379 & 0.6351 \\
    \multicolumn{1}{c|}{} & 64 & 0.7057 & 0.6375 & 0.6349 \\
    \hline
    \multicolumn{1}{c|}{\multirow{4}{*}{few-shot}} & 512 & \textbf{0.7927} & \textbf{0.5475} & \textbf{0.7196} \\
    \multicolumn{1}{c|}{} & 256 & 0.7917 & 0.5476 & 0.7098 \\
    \multicolumn{1}{c|}{} & 128 & 0.7897 & 0.5606 & 0.7099 \\
    \multicolumn{1}{c|}{} & 64 & 0.7901 & 0.5629 & 0.7099 \\
    \hline  
    \bottomrule          
\end{tabular}
}
\end{table}

\subsubsection{Impact of Distance Metric}

We empirically choose the cosine distance as the default choice for ReLLa to measure the semantic relevance, as it has been widely used to evaluate textual similarity in the realm of natural language processing (NLP)~\cite{wang2022text,reimers2019sentence,ni2021sentence}.
To offer a deeper insight, we compare three different distance metrics: (1) cosine distance, (2) L2 distance, and (3) L1 distance. 
We report the performance of ReLLa w.r.t. different distance metrics in both zero-shot and few-shot (<1\%) settings on MovieLens-1M dataset.
The results are given in Table~\ref{tab:ablation distance}. We have the following discussions:
\begin{itemize}[leftmargin=10pt]
    \item Cosine similarity inherently normalizes the vectors. It focuses on the angular difference between vectors rather than their magnitude. This means that even if two vectors differ greatly in size, they can still have a high cosine similarity if they point in similar directions.
    \item In high-dimensional spaces, L1 and L2 distances tend to suffer from the "curse of dimensionality", where the distance between all points becomes similar as dimensions increase. This makes these distances less effective measures of similarity in high dimensions. Cosine similarity is not affected by this issue and therefore remains effective in high-dimensional spaces.
\end{itemize}

\begin{table}[t]
    \caption{Ablation study w.r.t different distance metrics for ReLLa on MovieLens-1M dataset under both zero-shot and few-shot (<1\%) settings.
    }
    \vspace{-5pt}
    \label{tab:ablation distance}
    \resizebox{0.435\textwidth}{!}{
    \renewcommand\arraystretch{1.1}
    \begin{tabular}{c|c|ccc}
    \toprule
    \hline
    \multicolumn{1}{c|}{\multirow{2}{*}{Setting}} & \multicolumn{1}{c|}{\multirow{2}{*}{Distance}} & \multicolumn{3}{c}{MovieLens-1M} \\ 
    \multicolumn{1}{c|}{} & \multicolumn{1}{c|}{} & AUC & Log Loss & ACC \\
    \hline
    \multicolumn{1}{c|}{\multirow{3}{*}{zero-shot}} & Cosine & \textbf{0.7013} & \textbf{0.6250} & \textbf{0.6507} \\
    \multicolumn{1}{c|}{} & L2 & 0.6975 & 0.6356 & 0.6386 \\
    \multicolumn{1}{c|}{} & L1 & 0.6811 & 0.6388 & 0.6339 \\
    \hline
    \multicolumn{1}{c|}{\multirow{3}{*}{Few-shot}} & Cosine & \textbf{0.7927} & \textbf{0.5475} & \textbf{0.7196} \\
    \multicolumn{1}{c|}{} & L2 & 0.7872 & 0.5762 & 0.6944 \\
    \multicolumn{1}{c|}{} & L1 & 0.7833 & 0.5598 & 0.7119 \\
    \hline  
    \bottomrule          
\end{tabular}
}
\end{table}

\subsection{Potential Reason for Incomprehension}

We give our in-depth analysis and conjecture about the potential reason for the incomprehension problem of LLMs in recommendation. 
We argue that the incomprehension problem arises from the fact that LLMs struggle to comprehend highly heterogeneous user behavior sequences and thus fail to extract meaningful information from them. 
The heterogeneity of a sequence can be defined as the diversity of user behaviors within that sequence, such as different item genres.
The longer the user's behavior sequence, the higher the probability of it exhibiting a high level of heterogeneity, and consequently, the greater the difficulty for LLMs to comprehend.
Based on the conjecture above, retrieval, as the core technique for ReLLa, is essentially a means of homogenizing the user sequence. 
It aggregates homogeneous behaviors that are similar to the target item into a certain sequence and removes those unrelated heterogeneous behaviors, thus improving the comprehension capability of LLMs for long user sequences.

We provide an empirical study on MovieLens-1M dataset to demonstrate the homogenizing effect of retrieval. 
Here, we define the heterogeneity score as the number of unique movie genres in a given sequence of behaviors (i.e., movies). We illustrate two examples as follows:
\begin{itemize}[leftmargin=10pt]
    \item $[\text{Fiction, Comedy, Comedy, Family}]$ $\rightarrow$ heterogeneity score=3
    \item $[\text{Fiction, Fiction, Child, Fiction}]$ $\rightarrow$ heterogeneity score=2
\end{itemize}
In Table~\ref{tab:heter score}, we report the averaged heterogeneity score of two sequence types w.r.t. different length $K$: (1) the sequence is constructed by top-recent behaviors. (2) the sequence consists of top-relevant behaviors generated by our proposed semantic user behavior retrieval (SUBR).
From Table~\ref{tab:heter score}, we give the following discussions:
\begin{itemize}[leftmargin=10pt]
    \item We can see that the heterogeneity score gradually increases as the length $K$ grows. We argue that LLMs might fail to comprehend the given sequence once the heterogeneity score exceeds a certain threshold, and therefore we observe the phenomenon that the performance of LLM only peaks at around $K$=15 as illustrated in Figure~\ref{fig:illustration long context problem}.
    \item When equipped with our proposed SUBR, the heterogeneity score of top-relevant sequences largely decreases compared with the top-recent sequences. The lower the heterogeneity level of the sequence, the easier it is for LLMs to comprehend, and consequently, the better performance ReLLa can achieve with a larger $K$.
\end{itemize}

\begin{table}[t]
    \caption{The averaged heterogeneity scores of two sequence types w.r.t. different length $K$.
    }
    \label{tab:heter score}
    \resizebox{0.48\textwidth}{!}{
    \renewcommand\arraystretch{1.1}
    \begin{tabular}{c|ccccccc}
    \toprule
    \hline
    \multicolumn{1}{c|}{\multirow{2}{*}{Seq. Type}} & \multicolumn{6}{c}{MovieLens-1M} \\ 
    \multicolumn{1}{c|}{} & K=5 & K=10 & K=15 & K=20 & K=25 & K=30 \\ 
   \hline 
   Top Recent (Origin) & 2.91 & 4.19 & 5.09 & 5.80 & 6.39 & 6.90 \\ 
   Top Relevant (Retrieval) & 2.44 & 3.37 & 4.01 & 4.51 & 4.94 & 5.32 \\
   \hline  
   \bottomrule          
\end{tabular}
}
\end{table}

\end{document}